\documentclass{article}
\usepackage{jinstpub}
\usepackage[utf8]{inputenc}

\usepackage{amsmath,amssymb}
\usepackage{mathrsfs}
\usepackage[colorlinks=true]{hyperref}
\usepackage{graphicx}
\usepackage{lineno}
\usepackage{caption}
\usepackage{subcaption}

\title{Position Reconstruction in the DEAP-3600 Dark Matter Search Experiment}
\author[f]{P.~Adhikari,} 
\author[f,ab]{R.~Ajaj,}
\author[o]{M.~Alpízar-Venegas,}
\author[z]{P.-A.~Amaudruz,}
\author[f,ab]{J.~Anstey,}
\author[aa]{G.R.~Araujo,}
\author[a]{D.J.~Auty,}
\author[w]{M.~Baldwin,}
\author[m]{M.~Batygov,}
\author[a]{B.~Beltran,}
% ...

\affiliation[a]{Department of Physics, University of Alberta, Edmonton, Alberta, T6G 2R3, Canada}
% ...
\affiliation[f]{Department of Physics, Carleton University, Ottawa, Ontario, K1S 5B6, Canada}
% ...
\affiliation[m]{School of Natural Sciences, Laurentian University, Sudbury, Ontario, P3E 2C6, Canada}
% ...
\affiliation[o]{Instituto de Física, Universidad Nacional Autónoma de México, A. P. 20-364, Ciudad de México 01000, Mexico}
% ...
\affiliation[w]{Rutherford Appleton Laboratory, Harwell Oxford, Didcot OX11 0QX, United Kingdom}
% ...
\affiliation[z]{TRIUMF, Vancouver, British Columbia, V6T 2A3, Canada}
\affiliation[aa]{Department of Physics, Technische Universit\"at M\"unchen, 80333 Munich, Germany}
\affiliation[ab]{Arthur B. McDonald Canadian Astroparticle Physics Research Institute, Queen’s University, Kingston, ON, K7L 3N6, Canada}

\date{\today}

\emailAdd{deap-papers@snolab.ca}

\abstract{
In the DEAP-3600 dark matter search experiment, precise reconstruction of the positions of scattering events in liquid argon 
is key for background rejection and defining a fiducial volume that enhances dark matter candidate events identification. This paper describes three distinct position reconstruction algorithms employed by DEAP-3600, leveraging the spatial and temporal information provided by photomultipliers surrounding a spherical liquid argon vessel.
Two of these methods are maximum-likelihood algorithms: the first uses the spatial distribution of detected photoelectrons, while the second incorporates timing information from the detected scintillation light. Additionally, a machine learning approach based on the pattern of photoelectron counts across the photomultipliers is explored.
}


\begin{document}

\maketitle
 
\section{Introduction}
A compelling array of astrophysical observations indicates that a form of non-luminous dark matter not described by current physical theories 
comprises about 27\% of the energy density of the Universe \cite{planck}. By comparison, it appears that only about 5\% of the energy density 
is constituted by baryonic matter
described by the Standard Model of particle physics \cite{SM1,SM2}. 
Weakly interacting massive particles (WIMPs) are well-motivated dark matter candidates. Direct detection experiments try to observe the scattering of WIMPs 
with nuclei in massive detectors, located underground to shield them from cosmic backgrounds. 
The experimental WIMP-nucleus scattering signature is a low-energy deposit ($\lesssim$100 keV) caused by a nuclear recoil. 
Several experiments employing different detection media have set increasingly stringent limits on the existence of WIMPs \cite{WIMPS}. 

The DEAP-3600 experiment \cite{DEAPdetector}, located 2~km underground at SNOLAB, utilizes liquid argon (LAr) as dark matter detection medium. 
So far, DEAP-3600 has collected data using a LAr mass of \mbox{$(3269 \pm 24)$~kg}~\cite{DEAPspecificactivity39Ar}. The latest result from DEAP-3600 on WIMP searches is based on a \mbox{758~tonne$\cdot$day} exposure collected in 231~live-days \cite{DEAPdm2019}. 

\begin{figure}[htbp]
    \centering
    \includegraphics[width=0.5\textwidth]{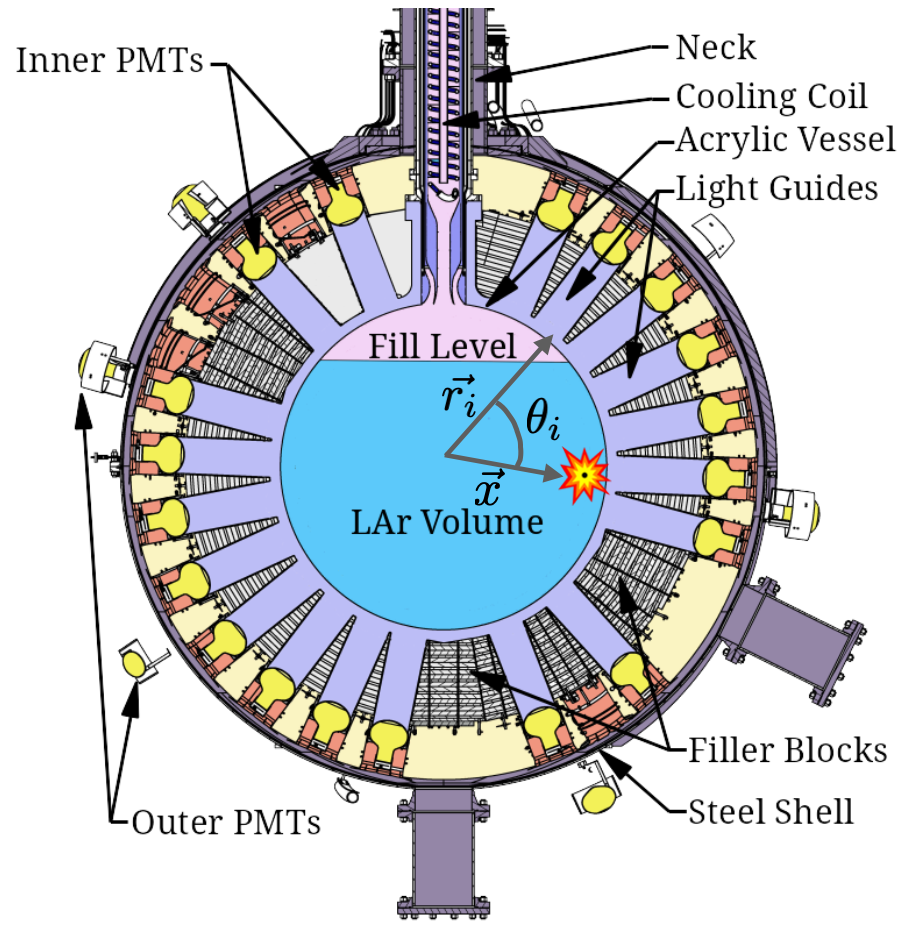}
    \caption{Schematic view of the DEAP-3600 detector (The water tank in which it is immersed is not shown). Variables used in the charge-based hit-pattern position reconstruction ($\vec{x}$, $\vec{r_i}$, $\theta_i$) are indicated.}
    \label{fig:DEAP}
\end{figure}

A detailed description of the DEAP-3600 detector can be found in \cite{DEAPdetector}, and a summary with the relevant information is reported here. 
In Figure~\ref{fig:DEAP}, a schematic design of the detector with its most relevant parts is shown. 
The detector’s main component is a \mbox{5~cm} thick spherical acrylic vessel (AV) with an inner radius of \mbox{85~cm} that can contain up to 3600 kg of LAr. 
The LAr is viewed by 255 Hamamatsu R5912-HQE high quantum efficiency photomultiplier tubes (PMTs) \cite{DEAPpmt} coupled with 45~cm long, 19~cm diameter acrylic light guides to the AV. The light guides provide shielding against neutrons to the LAr. Between the light guides, filler blocks made of high-density polyethylene and polystyrene contribute to neutron shielding. The inner surface of the AV is coated with a \mbox{$3\mu\rm{m}$} thick layer of tetraphenyl butadiene (TPB), deposited in situ and acting as a wavelength shifter \cite{TPB}. 
The photon spectrum from argon scintillation is in the vacuum ultraviolet (VUV) region, peaking at \mbox{128~nm}. This wavelength cannot excite the first atomic argon state, permitting light to travel through the LAr without absorption. When VUV photons reach the inner surface of the AV, they are absorbed by the TPB and re-emitted in the visible spectrum, peaking at around \mbox{420~nm}. This wavelength allows transmission through the acrylic and
is matched by the peak quantum efficiency of the PMTs.

LAr has unique scintillation time characteristics that allow the use of pulse-shape discrimination (PSD) \cite{DEAPpulseshape, DEAPscint} for distinguishing electromagnetic interactions, often called electronic recoils (ER), from nuclear recoil events (NR). 
PSD allows the suppression of ER events by a factor better than \mbox{$7.5\times 10^{-9}$} \cite{DEAPpulseshape} above the threshold used for the WIMP search. 
Events originating on the surface of the AV that contains the LAr are primarily due to $\alpha$ decays. These surface-based background events can be effectively rejected by using position reconstruction techniques that define a fiducial volume smaller than the full LAr content, thus excluding regions near the AV walls.

The AV is enclosed in a steel shell, which is immersed in a water tank acting as a veto detector. On the steel shell, 48 outward-facing PMTs are mounted for monitoring the water and vetoing cosmogenically-induced backgrounds.

The inner detector volume is accessed through a steel and acrylic neck connecting to the AV (see Figure.~\ref{fig:DEAP}). The neck contains a cooling coil that uses liquid nitrogen to cool the gaseous argon. 
Radioactivity from $\alpha$ particles on the surfaces of the neck can create scintillation events.
Significant shadowing of such events can lead to events reconstructed with a low energy that mimic WIMP recoils if they are mis-reconstructed into the main LAr volume. The presence of surface backgrounds in the AV and in the neck requires robust and precise methods for reconstructing the position of scintillation events.

This paper provides a detailed description of the position reconstruction algorithms employed by DEAP-3600 for volume fiducialization and background rejection. After a brief description of the photon counting and time information, three position reconstruction strategies are described and evaluated.
The first two methods employ a maximum-likelihood technique: one method is 
based on the ``hit pattern'' of the number of photoelectrons (PE) detected by each PMT in an event, while the other method is based on photons time-of-flight information.
The third method uses the hit pattern information as input to a neural network and extends position reconstruction to the detector's neck region.

\section{Photon Counting and Time Information} 
\label{sec:chargetime}
The position reconstruction algorithms are based on the number and arrival times of the PEs detected by each PMT, extracted from the digitized pulse shapes by custom algorithms \cite{DEAPpulseshape,qBayes}.
The reconstruction exploits an accurate Monte Carlo (MC) simulation based on the software {\tt RAT} \cite{RAT}, which combines {\tt Geant4} \cite{Geant4} simulation and analysis software {\tt ROOT} \cite{ROOT}, with event-based analysis tasks.
These algorithms maximally use the information measured by the PMTs, which is key for an accurate position reconstruction. 

The DEAP-3600 data acquisition system records waveforms of the PMTs from \mbox{-2.6~$\mu$s} to \mbox{14.4~$\mu$s} relative to the trigger time using CAEN V1720 \mbox{250~MHz} digitizers \cite{CAEN_V1720}. The raw pulses from the PMTs have a rise time of about \mbox{2~ns}, and are thus slowed down and amplified using signal conditioning boards \cite{DEAPdetector} to allow the digitizers to optimally determine both pulse height and timing. This results in single PE pulses at the digitizer level that are about \mbox{20~ns} wide and \mbox{20 mV} in amplitude. Each PMT waveform has its baseline removed by subtracting an offset calculated by averaging 62 baseline samples. 
The photon count done in this way is biased because the PMTs have significant afterpulsing, which manifests as additional pulses that don't correspond to scintillation photons. 

A low-bias estimate of the photon count can be obtained using a likelihood fit~\cite{DEAPpulseshape}. 
With this method, the pulse is modelled as  $n_{\rm PE}$ photoelectrons, which is the sum of 
signals from scintillation photons and 
signals from afterpulsing.  
Bayes' theorem is used to estimate $n_{\rm PE}$ at the time of each pulse, given the pulse charge, the liquid argon scintillation probability density function, the times of preceding pulses, the afterpulsing time and charge probability density function, and the single PE charge distribution of the PMT \cite{Butcher}.
This PE estimator is used as input to both the hit pattern algorithm and the machine-learning algorithm.

\section{Position Reconstruction Algorithms}
In DEAP-3600, two maximum-likelihood algorithms are used to reconstruct the interaction vertex position. Both approaches fit a model where a flash of VUV scintillation light is emitted from a point within the spherical inner volume of the detector and is propagated to the wavelength shifter and the PMTs.  
The first algorithm is based on the spatial PMT hit patterns which are analyzed to derive the most likely position of the interaction vertex. 
The second approach is based on the photons' time-of-flight and leverages the finite speed of light in LAr: the general idea is that PMTs closer to the source produce signals earlier than distant ones. 

A third approach uses a supervised machine-learning algorithm based on the spatial hit pattern, with the flexibility to train the method on specific regions of the detector and on different event types. 

\subsection{Hit Pattern Algorithm}
\label{sec:mbl}
This method is based on the pattern of charges measured by the 255 PMTs around the AV. 
The position is calculated by maximizing a likelihood function, constructed with a probability distribution
determined with a MC simulation of the detector.

For a given simulated event position, the number of photons measured by each of the PMTs is calculated with the full optical model in our MC simulation \cite{OpticalModel} which includes the refractive index, group velocity, and Rayleigh scattering
length for VUV and visible photons as well as their dependence on thermodynamic properties.
The simulation also takes into account the solid angle of the PMTs as viewed from the vertex, light absorption, 
re-emission at the TPB wavelength shifter, diffusion, and reflections at surfaces.
The probability of detecting photons by a PMT is a highly non-linear function of the event's coordinates and energy, which changes rapidly as the interaction point comes closer to the PMT and has to be determined with the simulation.  
To determine the event's location $\vec{x}$, the following likelihood is maximized with a simplex algorithm:
\begin{equation}
    \ln \mathcal{L}(\vec{x}) = \sum^{\rm{N_{PMTs}}}_{i=1} \ln \left[ \rm{P}(n_i; \, \lambda_i)\right] \,,
    \label{eq:PE_likelihood}
\end{equation}
where $P(n_i;\lambda_i)$ is the Poisson probability of observing $n_i$ PE in PMT~$i$ over the full \mbox{10~$\mu$s} time window. The coordinate system has its origin $\vec{x}=(0,0,0)$ at the center of the spherical AV 
and the $z$ axis pointing in the vertical direction towards the detector's neck.
The parameter 
\begin{equation}
\lambda_i\left(|\vec{x}|, \theta_i,n_{\rm PE} \right) = n_{\rm PE} \lambda_i\left(|\vec{x}|, \theta_i \right) \,,
\end{equation}
describes the expected number of PE hitting the PMT $i$ located at position $\vec{r}_i$,
given the event position $\vec{x}$, an angle $\theta_i$ between $\vec{x}$ and $\vec{r}_i$,
and the total number of detected 
photons $n_{\rm PE}$ (see Figure~\ref{fig:DEAP} for a visual description of the variables).
The factorized total number of photons $n_{\rm PE}$ provides an absolute normalization for $\lambda_i$.
The simulation assumes a completely filled detector and the simulated events are generated on a grid with 20 points along the positive $x$ axis, 20 along the positive $y$ axis, and 40 symmetrically on the positive and negative $z$ axis. This choice results in 20 values for the radius $r=\sqrt{x^2+y^2+z^2}$.  
Assuming the cylindrical symmetry of the detector, events with $x<0$ and $y<0$ are modeled as the corresponding positive coordinate values. 
The $z$ direction is probed in both directions to account for the asymmetry introduced by the 
presence of the detector neck. 
The points are chosen to minimize their total number while probing the volume uniformly. Since these points lay on the three axes, the influence of individual direction-dependent effects can be tested.
After the simulation, we define 20 cubic spline interpolated functions $\lambda_i(\cos\theta;r_j/r_0) $ with the index $j$ corresponding to each of the radius values.  Here $r=|\vec{x}|$ and $r_0=85$~cm is the inner radius of the vessel.  For a given angle, a second cubic spline $\lambda(r/r_0;\cos\theta)$ is defined to interpolate in the radius variable.    







\subsection{Time-of-Flight Algorithm}
\begin{figure}[htbp]
    \centering
    \includegraphics[scale=0.3]{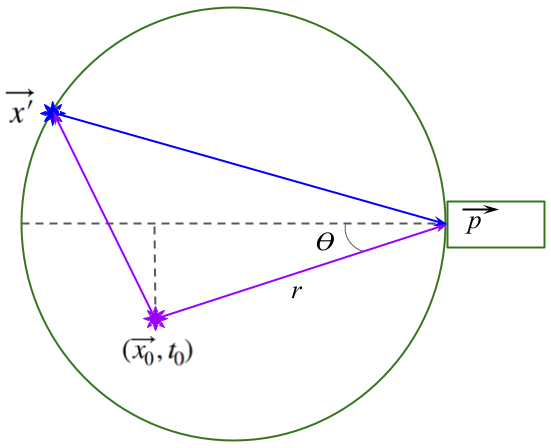}    
    \caption{The light propagation model and the coordinate system for the time-of-flight reconstruction algorithm.}
    \label{fig:TOF_schematic}
\end{figure}

The time-of-flight algorithm for position reconstruction calculates the arrival time of a photon at each PMT.
The algorithm uses a simplified optical model, illustrated in Figure~\ref{fig:TOF_schematic}.
For a scintillation interaction event taking place at event time $t_0$ and position $\vec{x_0}$ in LAr, the arrival time of a photon at a PMT can be calculated as
\begin{equation}
    t_i = t_0 + T_f(\Vec{p},\Vec{x_0}) + \tau \,,
    \label{eq:t_arrival}
\end{equation}
in which $T_f(\Vec{p},\Vec{x_0})$ is the time-of-flight from the event vertex $\Vec{x_0}$ to position $\Vec{p}$, the center of the light guide front surface. The time $\tau$ is a random variable based on three physical processes: singlet decay of argon with a time constant of \mbox{2.2~ns}, TPB re-emission with a time constant of \mbox{1.5~ns}, and a resolution term of \mbox{0.6~ns} that includes PMT response, resolution,  and variability in the propagation time through the light guides. 
These effective values for the time constants optimally match data with simulation.
The vector $\Vec{p}$ rather than the physical PMT position is used both for the determination of time calibration constants and for reconstruction.
%
When a VUV photon is wavelength-shifted by TPB at a point $\Vec{x'}$ on the acrylic surface, the resulting visible photon can be emitted into a light guide (in which case $\Vec{p} = \Vec{x'}$) or be emitted back into the inner detector volume and travel with a greater group velocity before reaching a light guide front point $\Vec{p}$. The time-of-flight can then be calculated,
\begin{equation}
    T_f(\Vec{p},\Vec{x_0}) = \frac{|\Vec{x'} - \Vec{x_0} |}{\rm{v_{uv}}} + \frac{|\Vec{p} - \Vec{x'}|}{\rm{v_{vis}}} \,,
    \label{eq:TOF-eq}
\end{equation}
where $\rm{v_{uv}}$ is the group velocity \mbox{(133~mm/ns)}
of the \mbox{128~nm} VUV light~\cite{GroupVelLAr} and $\rm{v_{vis}}$ is the group velocity (\mbox{241~mm/ns}) of \mbox{420~nm} visible light in LAr~\cite{GroupVelVis}. 
A test with the values of group velocities shifted by $\pm5\%$ resulted in a negligible effect on the position reconstruction resolution. Reflections and scattering of visible photons in TPB and Rayleigh scattering in the LAr are neglected in this algorithm.

$P_t(t;r,\sin\theta)$ is the probability density function (PDF) that corresponds to the flight time from a particular position, with $r$ and $\sin\theta$  defined in Figure~\ref{fig:TOF_schematic}. 
The probability of measuring a photon in a PMT-light guide assembly at position $\Vec{p}$ starting from a position $\Vec{x_{0}}$ between times $t_{1}$ and $t_{2}$ is described by the following equations:
\begin{equation}\label{beginFirstBlock}
P_t(t_{1},t_{2};\Vec{p},\Vec{x_{0}})= \frac{1}{(4\pi)^{2}}\intop_{-1}^{1}d\cos\theta_{\mathrm{uv}}\intop_{0}^{2\pi}d\phi_{\mathrm{uv}}\intop_{-1}^{1}d\cos\theta_{\mathrm{vis}}\intop_{0}^{2\pi}d\phi_{\mathrm{vis}}H\left(t_{1,}t_{2},\frac{d_{\mathrm{uv}}}{\rm{v_{uv}}}+\frac{d_{\mathrm{vis}}}{\rm{v_{vis}}}\right)G(\theta_{\mathrm{vis}},\phi_{\mathrm{vis}})
\,,
\end{equation}
\begin{equation} \label{secondeq}
r=\left|\Vec{x_{0}}+d_{\mathrm{uv}}\left(\sin\theta_{\mathrm{uv}}\cos\phi_{\mathrm{uv}},\sin\theta_{\mathrm{uv}}\sin\phi_{\mathrm{uv}},\cos\theta_{\mathrm{uv}}\right)\right| \,,
\end{equation}
\begin{equation}\label{thirdeq}
d_{\mathrm{vis}}=\left|\Vec{x_{0}}+d_{\mathrm{uv}}\left(\sin\theta_{\mathrm{uv}}\cos\phi_{\mathrm{uv}},\sin\theta_{\mathrm{uv}}\sin\phi_{\mathrm{uv}},\cos\theta_{\mathrm{uv}}\right)-\Vec{p}\right| \,,
\end{equation}
\begin{equation}
H(t_{1},t_{2},x)=\begin{cases}
1 & t_{1}\leq x\leq t_{2}\,, \\
0 & \mathrm{otherwise}\,,
\end{cases}
\end{equation}
\begin{equation}
G(\theta_{\mathrm{vis}},\phi_{\mathrm{vis}})=
\begin{cases}
1 & \begin{aligned}r_{\mathrm{lg}} >  |\Vec{x_{0}} + & d_{\mathrm{uv}}\left(\sin\theta_{\mathrm{uv}}\cos\phi_{\mathrm{uv}},\sin\theta_{\mathrm{uv}}\sin\phi_{\mathrm{uv}},\cos\theta_{\mathrm{uv}}\right)\\
	+ & d'_{\mathrm{vis}}\left(\sin\theta_{\mathrm{vis}}\cos\phi_{\mathrm{vis}},\sin\theta_{\mathrm{vis}}\sin\phi_{\mathrm{vis}},\cos\theta_{\mathrm{vis}}\right)-\Vec{p}|\end{aligned}\,, \\
0 & \mathrm{otherwise} \,.
\end{cases}
\label{endFirstBlock}
\end{equation}
Eq.~\ref{beginFirstBlock} gives the overall probability $P_t$ and assumes that the scintillation light and the TPB fluorescence emission are isotropic. The functions $H$ and $G$ impose the conditions that the time-of-flight must be between $t_{1}$ and $t_{2}$ and that the light from the TPB fluorescence strikes the light guide of interest. 
$d_{\mathrm{uv}}$ and $d_{\mathrm{vis}}$ are the distances traveled by VUV and visible light, respectively.
$r_{\mathrm{lg}}=9.5\,\rm{cm}$ is the radius of the face of the light guide. 
Eq.~\ref{secondeq} is solved to determine $d_{\mathrm{uv}}$ and to impose the geometry of the TPB surface at a radius of $r=85$~cm, which marks the termination of the VUV light. 
Eq.~\ref{thirdeq} is the distance between the TPB and PMT. 
For the time-of-flight calculation, we approximate that the distance to the center of the light guide can always be calculated.
This is an approximation because the time-of-flight in the detector depends on where the photon strikes the face of the light guide. However, the largest uncertainty arises from fluorescence occurring directly in front of the PMT of interest, with a maximum value of approximately \mbox{1/3~ns} $\left(r_{\mathrm{lg}}/\mathrm{v_{vis}}\right)$.
This uncertainty is accounted for in the time resolution of the PMT-light guide system.

For the calculation of the solid angle $G$, we use the distance traveled by visible light between the TPB fluorescence and the acrylic vessel at the face of the light guide, $d'_{\mathrm{vis}}$, and check if the point of impact is within the radius $(r_{\mathrm{lg}})$ of the face of the light guide. 

The likelihood $\mathcal{L}(t_0,\Vec{x_0})$ for a given event time $t_0$ and position $\Vec{x_0}$ is
\begin{equation}
    \ln \mathcal{L}(t_0,\Vec{x_0}) = \sum_{i=1}^N \ln P_t(t_i-t_0; r_i, \sin\theta_i),
\label{eq:TOF_likelihood}
\end{equation}
where we sum over the $N$ pulses in the first 35 ns, with the detection time $t_i$, distance $r_i$, and angle $\theta_i$, based on the relevant PMT and vertex position.  A time-of-flight correction is applied if the path from vertex to PMT passes through the gas layer on the top of the detector.  The likelihood in Eq.~\ref{eq:TOF_likelihood} is maximized using the {\tt TMinuit2} fitter in ROOT~\cite{ROOT}.  

Since the calculation of the PDF ($P_t$) requires significant time, we employ interpolation techniques to evaluate the PDF while fitting. The detailed calculation is performed at discrete points $r=0,1,2,...,170~\rm{cm}$, and at $\sin\theta=0,0.01,0.02,...,1.0$, and in 0.25 ns time bins.  The PDFs are then convolved with a time distribution that accounts for the random variable $\tau$ from Eq.~\ref{eq:t_arrival}.  
We then use a cubic spline interpolation in time, followed by a bilinear interpolation in $r$ and $\sin\theta$ during the fit. 
\begin{figure}
    \centering
    \includegraphics[scale=0.38]{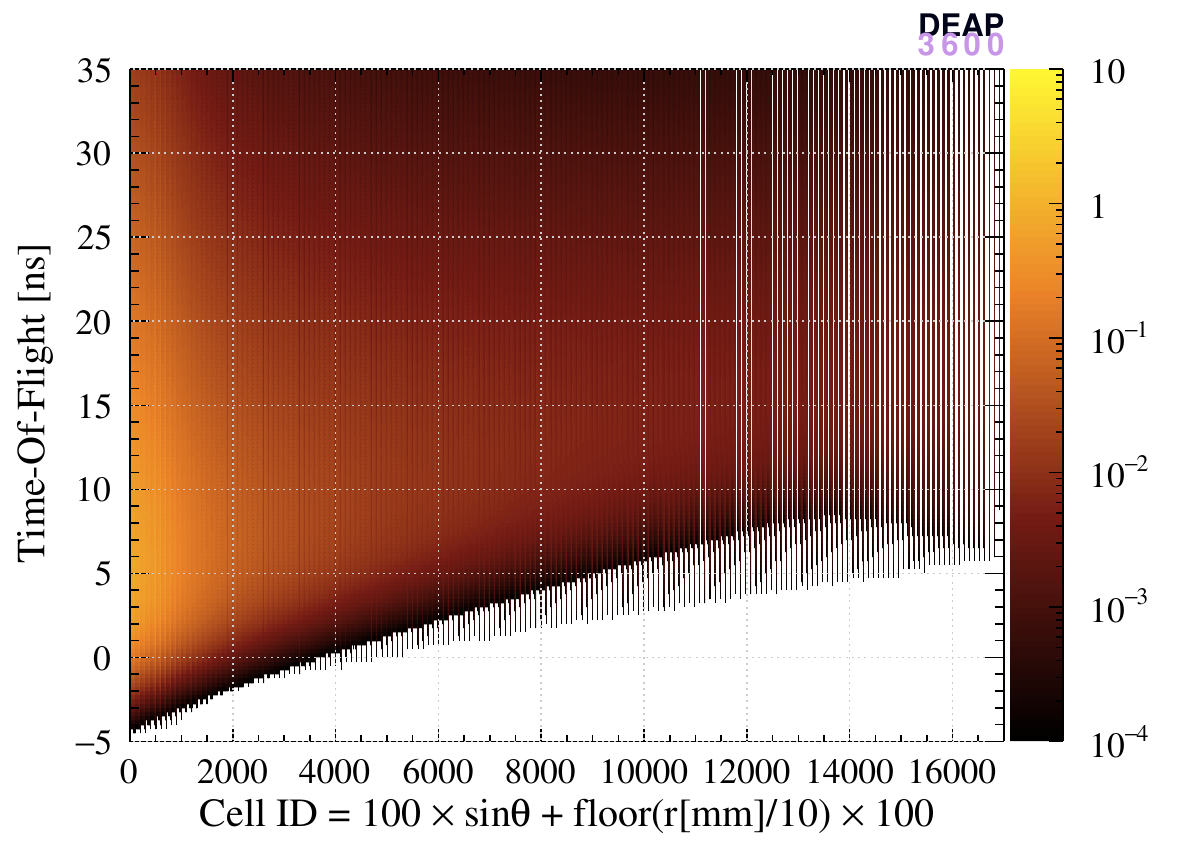}
    \includegraphics[scale=0.38]{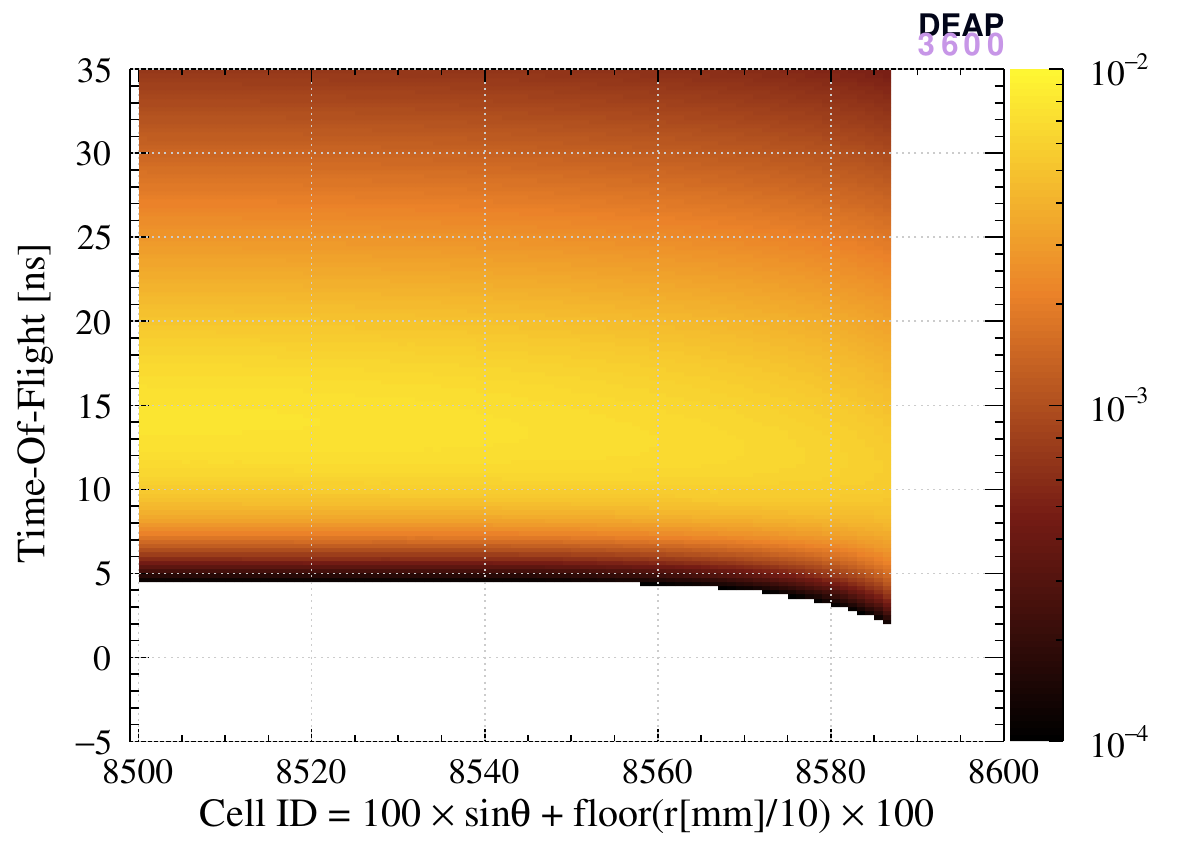}    
    \caption{The probability density function is shown as a two-dimensional histogram for the time-of-flight and spatial "cell" ID, with the color scale proportional to the probability density. The variables $r$ and $\theta$ in the definition of cell ID are given in Fig.~\ref{fig:TOF_schematic}. Left: The histogram includes all cells. Right: The histogram is zoomed to show cells 8400 to 8700, representing three distance values $r$ = 840~mm, 850~mm, and 860~mm with $\sin\theta$ varying from 0.00 to 0.99 for each distance value.}
    \label{fig:TF2PDF-dist}
\end{figure}
The left panel of Figure~\ref{fig:TF2PDF-dist} shows the convolved PDF in a two-dimensional representation where the vertical axis is the time-of-flight, the horizontal axis reports the identification numbers (ID) of spatial cells defined as ID~$=100\times \sin\theta +\textrm{\tt floor}(r[\textrm{mm}]/10)\times 100$ and the colour is the value of the PDF.
The right panel shows a zoomed view around $r$ = 850~mm representing a $\sin\theta$ range from 0.00 to 0.99.
For large values of $\sin\theta$, visible light arrives earlier than most VUV light (for cell numbers greater than 8560) because these cells are close to the surface. 

\subsection{Machine-Learning  Algorithm}

The non-linearity of the position reconstruction problem in LAr points towards a solution employing machine learning (ML). The good performance of the hit pattern algorithm-based likelihood reconstruction suggests a similar approach using a supervised training algorithm. Another strong motivation for considering a ML approach is the possibility of training it with scintillation events in the LAr volume and events in other parts of the detector. In this context, the ML training includes events originating in the neck of the detector which constitute a relevant background for dark matter searches. 

We investigated several algorithms, from {\em Naive Bayes} \cite{naivebayes} to more complex models based on boosted decision trees \cite{BDT} and neural networks. Our choice converged on a fully connected feed-forward neural network (FFNN) \cite{NN}
for its better performance, relative simplicity, and training speed \cite{Ilyasov}.

For the position reconstruction in the DEAP-3600 detector volume, the algorithm must realize a mapping between the 255 input charge values measured by the PMTs and the $(x, y, z)$ position coordinates as output. In practice, three single-output FFNN models were trained: one for each output Cartesian coordinate.
Including time-of-flight information as additional FFNN inputs was also tested, but no significant improvements in performance were observed. 
The final network architectures of the selected FFNNs is reported in Table~\ref{tab:NN}.
The number of layers and neurons was determined by iteratively enlarging the network
until no increase in the performance was observed.

\begin{table}[!ht]
    \centering
    \begin{tabular}{c|cc}
         \hline
         {\bf Coordinates $x, y$} & {\bf \# Neurons} & {\bf Activation Function} \\   
        \hline  
         Input Layer & 255 & ReLU \\
         Hidden Layer 1 & 255 & ReLU \\
         Hidden Layer 2 & 255 & ReLU \\
         Hidden Layer 3 & 16 & ReLU \\
         Output Layer & 1 & linear \\
         \hline
         {\bf Coordinate $z$} & & \\
         \hline
         Input Layer & 255 & ReLU \\
         Hidden Layer 1 & 1024 & ReLU \\
         Hidden Layer 2 & 1024 & ReLU \\
         Hidden Layer 3 & 1024 & ReLU \\
         Output Layer & 1 & linear \\
         \hline
    \end{tabular}
    \caption{Architecture of the three FFNNs used for position reconstruction. The coordinates $x$ and $y$ are reconstructed with the same FFNN topology. 
    ReLU is the rectifier linear unit function ${\rm ReLU}(x) = \max(0, x)$.}
    \label{tab:NN}
\end{table}

The training of the FFNNs was performed using MC simulations of $10^5$ $\beta$-decays from $^{39}$Ar, $10^5$ nuclear recoils on $^{40}$Ar, and $10^5$ shadowed $\alpha$-decays from $^{210}$Po happening on the acrylic argon flow-guides in the neck. 
These samples are divided into training and validation sets with proportions of $70:30$. 
The training process for each Cartesian coordinate minimizes the mean squared distance (MSD) between the predicted position $y_{\rm pred,i}$ and the true simulated position $y_{\rm true,i}$ for each event $i$:
\begin{equation}
    MSD = \frac{1}{N}\sum_{i=1}^{N}\left( y_{{\rm pred},i} - y_{{\rm true},i} \right)^2 \,,
\end{equation}
where $N$ is the number of input events presented to the neural network during the training stage.
Training for reconstructing the $x$ and $y$ coordinates required 400 epochs, while the $z$ coordinate  required 40 epochs.


\section{Performance of the Position Reconstruction Algorithms on Data and Simulations}

The DEAP-3600 experiment’s WIMP-search analysis primarily utilized the hit-pattern algorithm for fiducialization, with additional cross-validation against the time-of-flight algorithm to ensure consistency \cite{DEAPdm2019}. Both the hit pattern and time-of-flight algorithms apply a likelihood framework based on simulated events within the detector’s LAr region.

In contrast, the ML algorithm was specifically trained on interactions occurring in both the LAr region and the detector’s neck region. It was optimized to maximize discrimination between these two event locations. However, due to the limited availability of high-statistics samples of real neck events, the ML algorithm’s performance was evaluated in this case exclusively using simulated data, benchmarked against the hit-pattern and time-of-flight algorithms.

In the following, we compare the different algorithms in in their ability to reconstruct
the position of events in liquid argon. 



\subsection{Position reconstruction with $^{39}$Ar $\beta$-Decays and $^{40}$Ar Nuclear Recoils}
\label{sec:ar39}

To compare the three algorithms in their ability to reconstruct the position of scattering
events from $\beta$-decays of $^{39}$Ar and $^{40}$Ar recoils in LAr, we apply an appropriate selection to data and MC generated events.
We consider events that have 
(\romannumeral 1) good quality waveforms recorded by the digitizers, e.g. well-formed baseline voltage and valid trigger time,
(\romannumeral 2) physics trigger signals only, excluding calibration/cosmics triggers, 
(\romannumeral 3) no pulse multiplicity consistent with two or more coincident physics events.
To avoid light leakage from a previous event, the number of pulses registered in the first 1600~ns of the waveform must be
less than or equal to three 
and the time between the event and the previous one must be greater than \mbox{20~$\mu$}s.
The reconstructed number of PE is selected to be within 90--200 PE, matching the WIMP region of interest (ROI)~\cite{DEAPdm2019}.

\begin{figure}[htbp]
    \centering
    \includegraphics[scale=0.38]{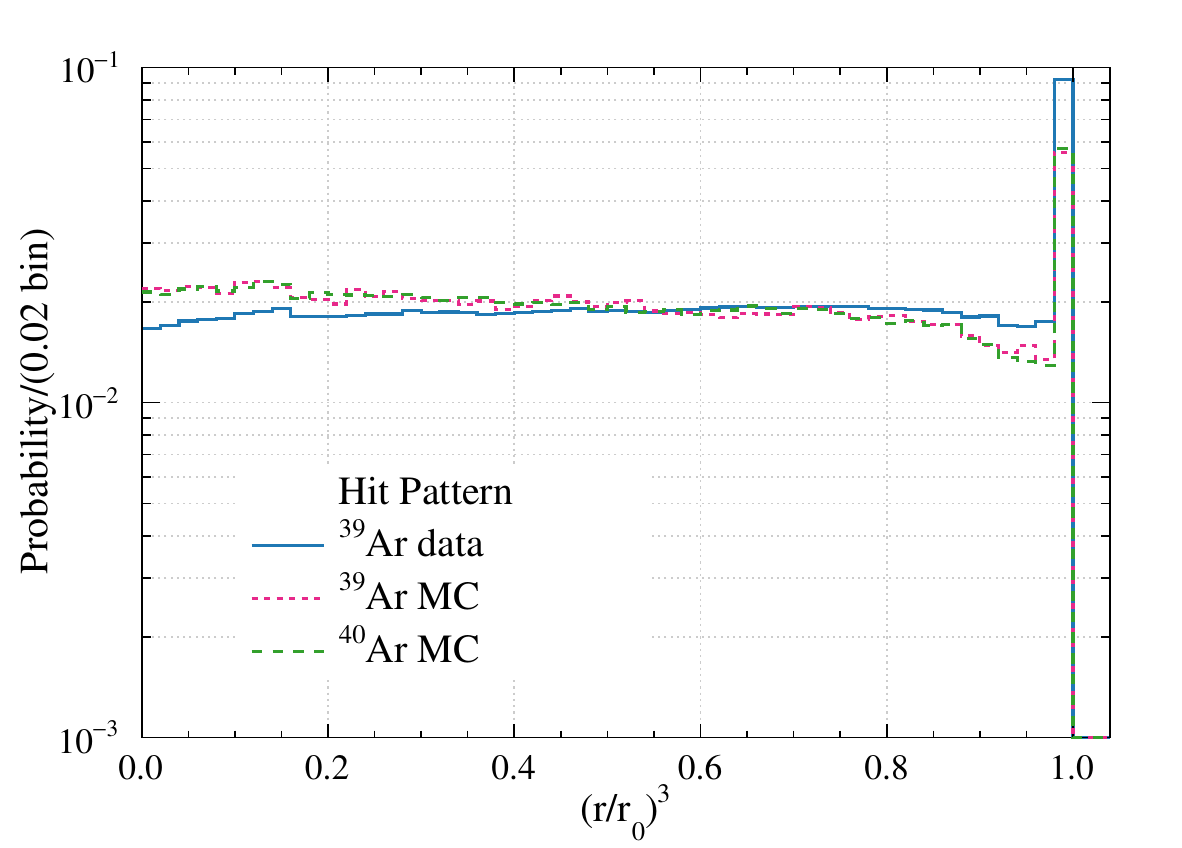}
    \includegraphics[scale=0.38]{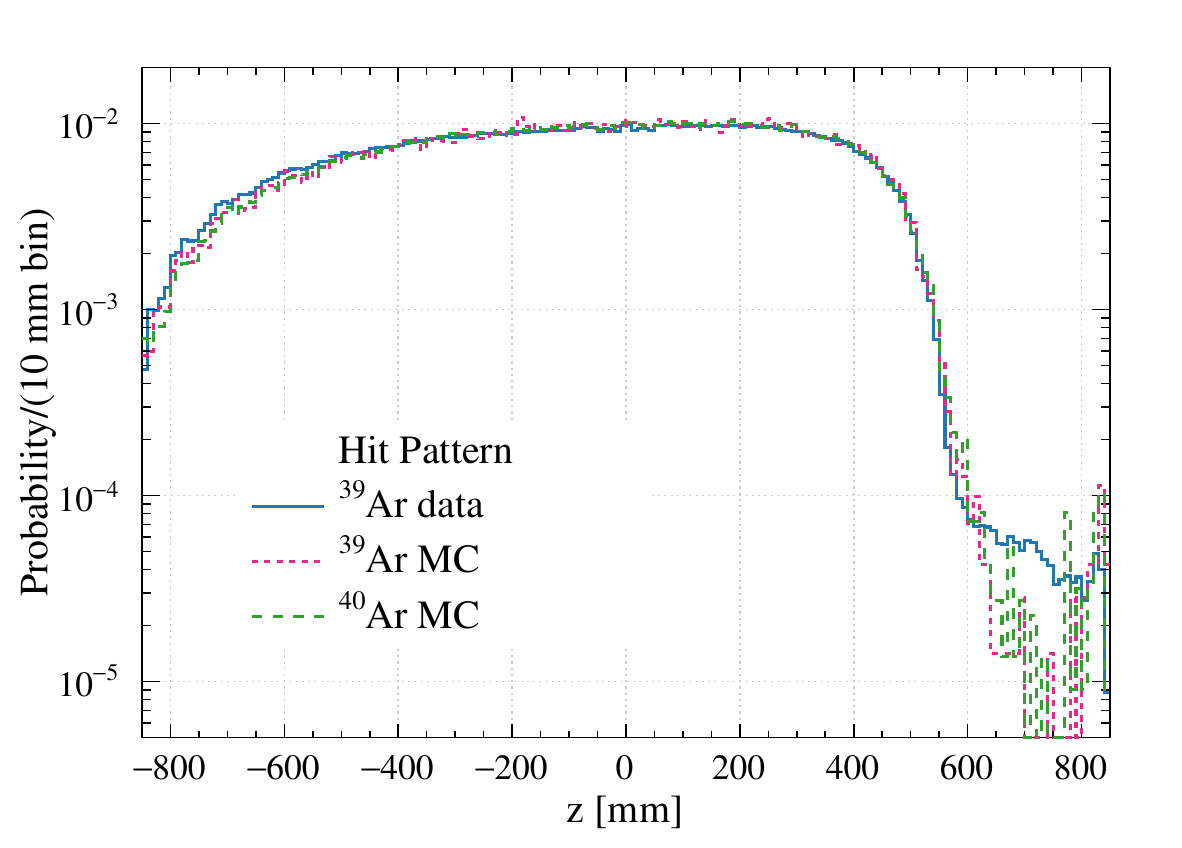}
    \includegraphics[scale=0.38]{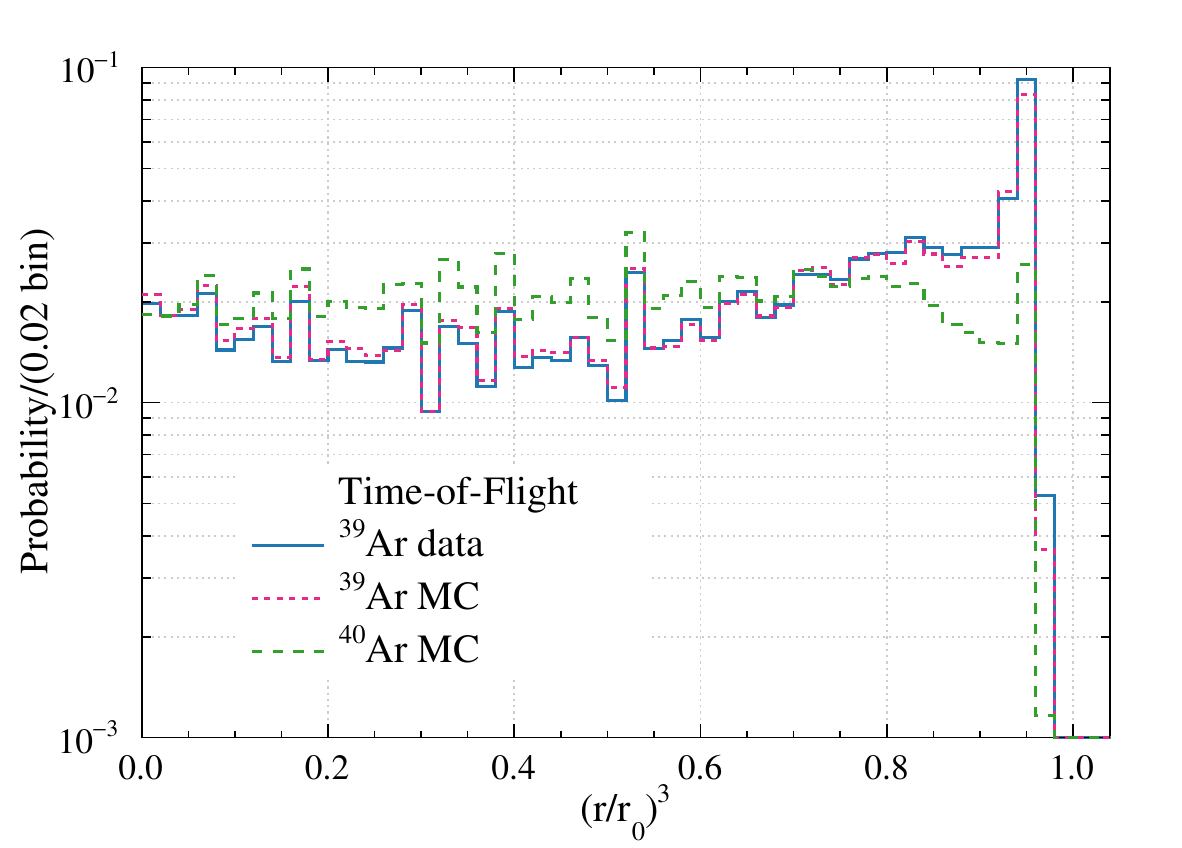}
    \includegraphics[scale=0.38]{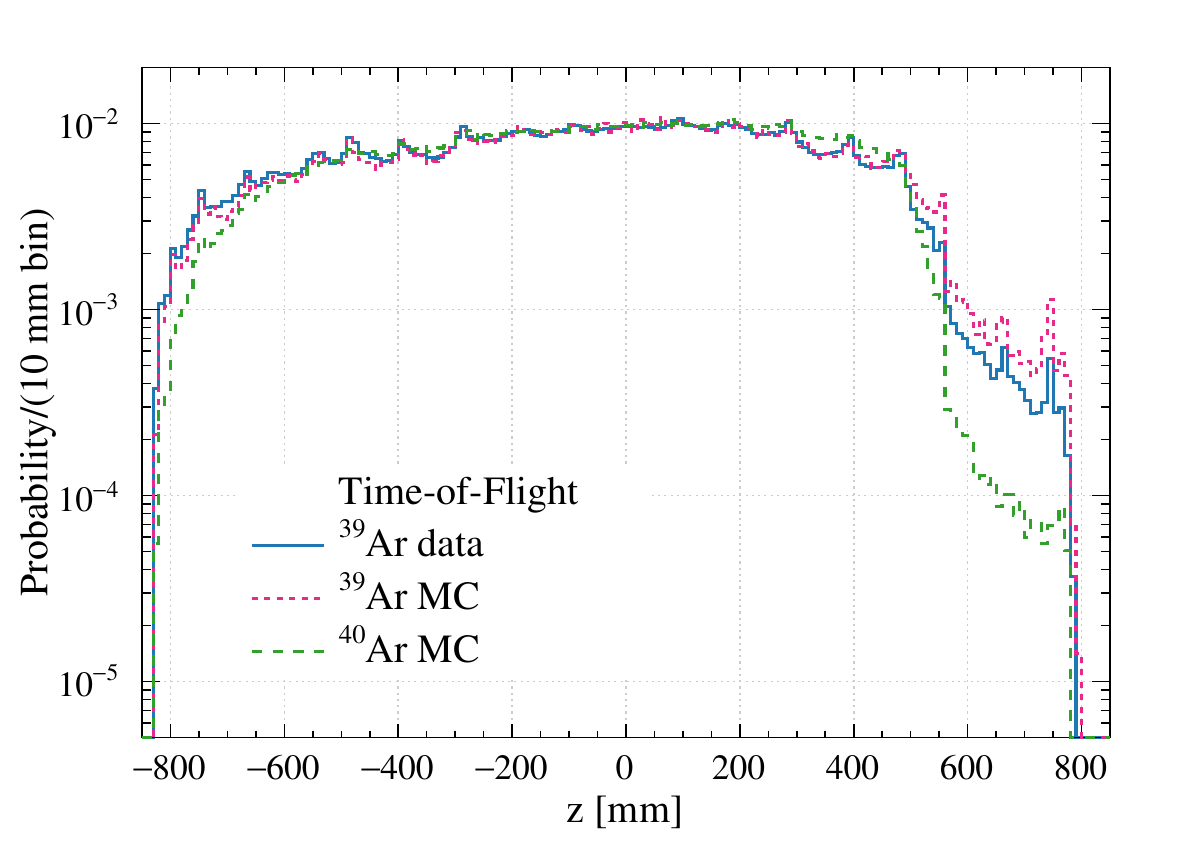}
    \includegraphics[scale=0.38]{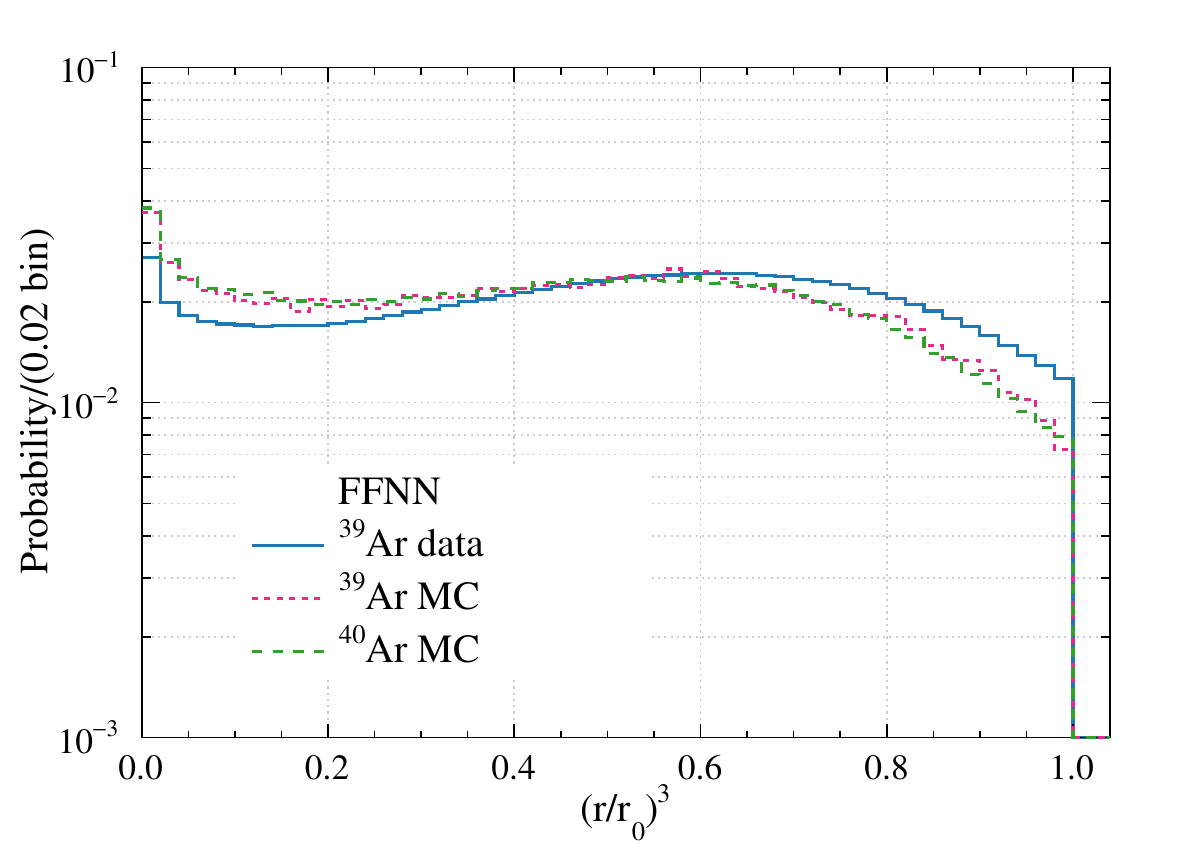}
    \includegraphics[scale=0.38]{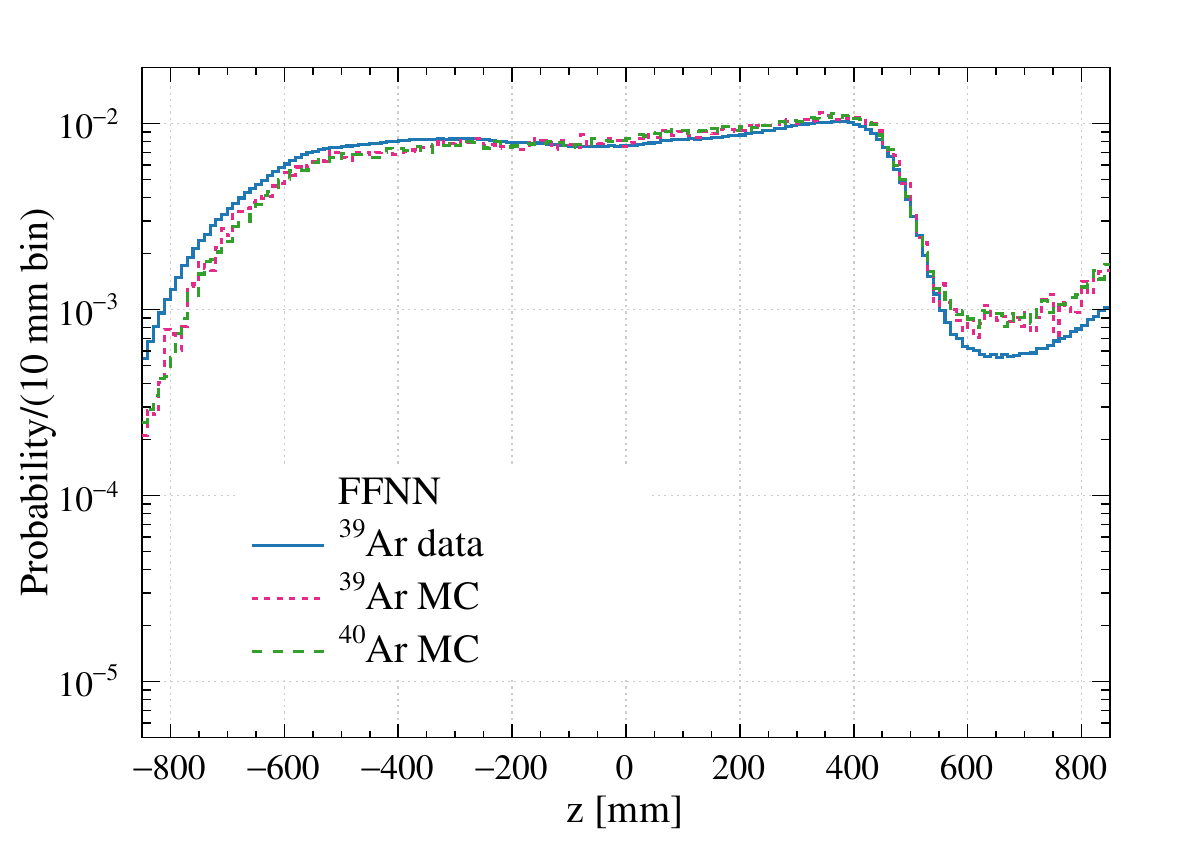}
    \caption{Distributions of reconstructed angular position variables, normalized reconstructed cubed radius ($(r/850\,\rm{mm})^3$) and reconstructed $z$, comparing $^{39}\rm{Ar}$ data (solid) and $^{39}\rm{Ar}$ simulation (dotted) and $^{40}\rm{Ar}$ nuclear recoil simulation (dashed), for the hit pattern algorithm (top row), the time-of-flight algorithm (middle row), and the neural network algorithm (bottom row). 
    All the plots are normalized with their respective area. 
    }
    \label{fig:RbyR0_cubed_MBL_TF2}
\end{figure}

\begin{figure}[htbp]
    \centering
    \includegraphics[scale=0.5]{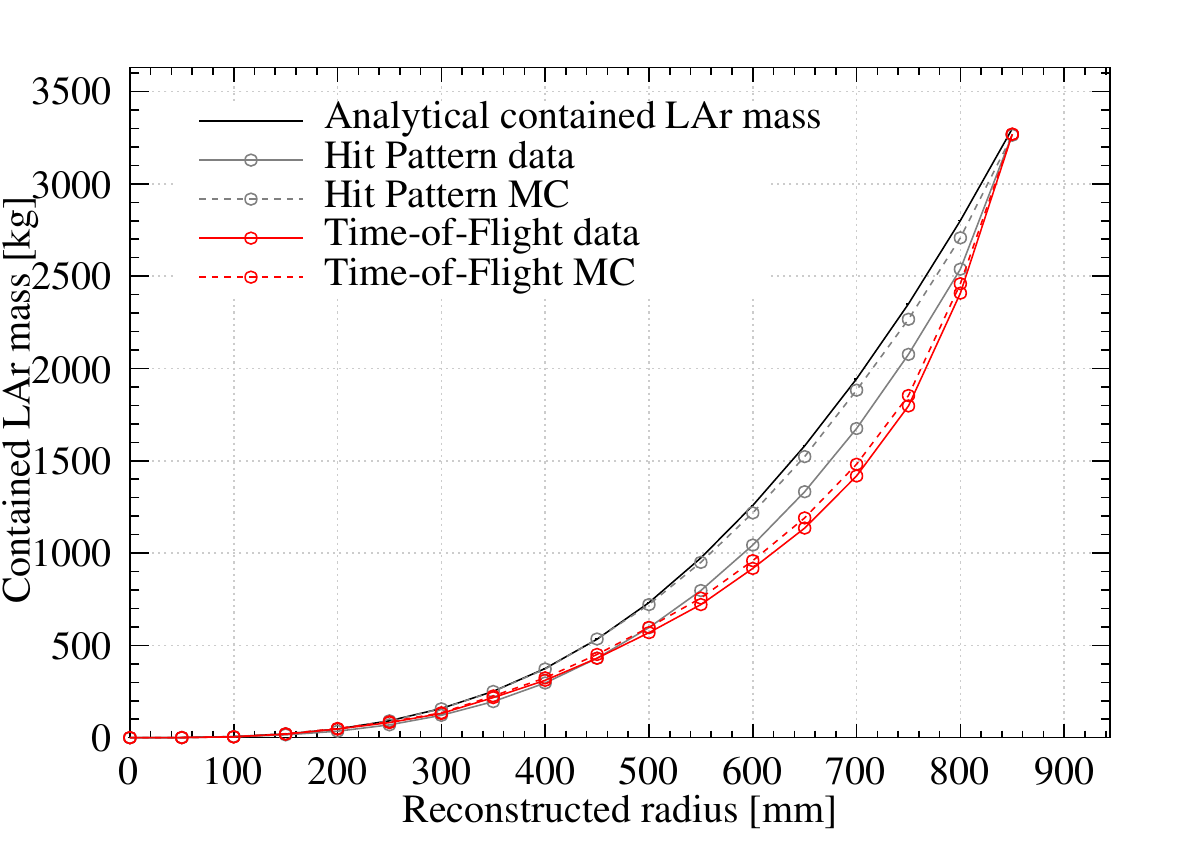}
    \caption{Estimates from the hit pattern and time-of-flight based algorithms of the contained mass of LAr within a radius of the reconstructed position. The estimate is based on 
    the fraction of $^{39}$Ar decays observed
    in the 90--200 PE range reconstructing within a given radius. It is assumed that the true positions of the $^{39}$Ar nuclei are uniformly distributed throughout the LAr target. 
    The analytical contained LAr mass calculated using the geometric volume and the density is also shown.
 }
    \label{fig:contained_LAr_mass}
\end{figure}

The left panels of Figure~\ref{fig:RbyR0_cubed_MBL_TF2} show the normalized cubed radius distribution of reconstructed $^{39}\rm{Ar}$ ER events (data and MC) and $^{40}\rm{Ar}$ nuclear recoil events (MC) with the three algorithms and 
the right panels show the $z$ distributions. 
The hit pattern algorithm produces a more uniform distribution in the bulk region and is biased near the AV surface. In contrast, the distribution obtained with the time-of-flight algorithm is less smooth in the bulk while less biased near the surface. 
The enhancement of the distribution is more pronounced for the hit pattern algorithm since the probability 
distribution used in the likelihood is strongly non-linear when an event is very close to a PMT which collects most of the light.
Figure~\ref{fig:contained_LAr_mass} shows the contained LAr mass for Hit Pattern and Time-of-Flight based algorithms for $^{39}$Ar data and MC along with the curve calculated from the geometric volume of the detector. 
All the algorithms are found to underestimate the contained LAr mass as function of the radius with respect to the simplified ideal geometric calculation. The hit pattern algorithm applied to simulated data achieves the best result, although, as visible also in Figure~\ref{fig:RbyR0_cubed_MBL_TF2} (top-left), it tends to reconstruct events closer to the LAr surface. This effect is more evident when real data are considered.

\begin{figure}[htbp]
    \centering
    \includegraphics[scale=0.38]{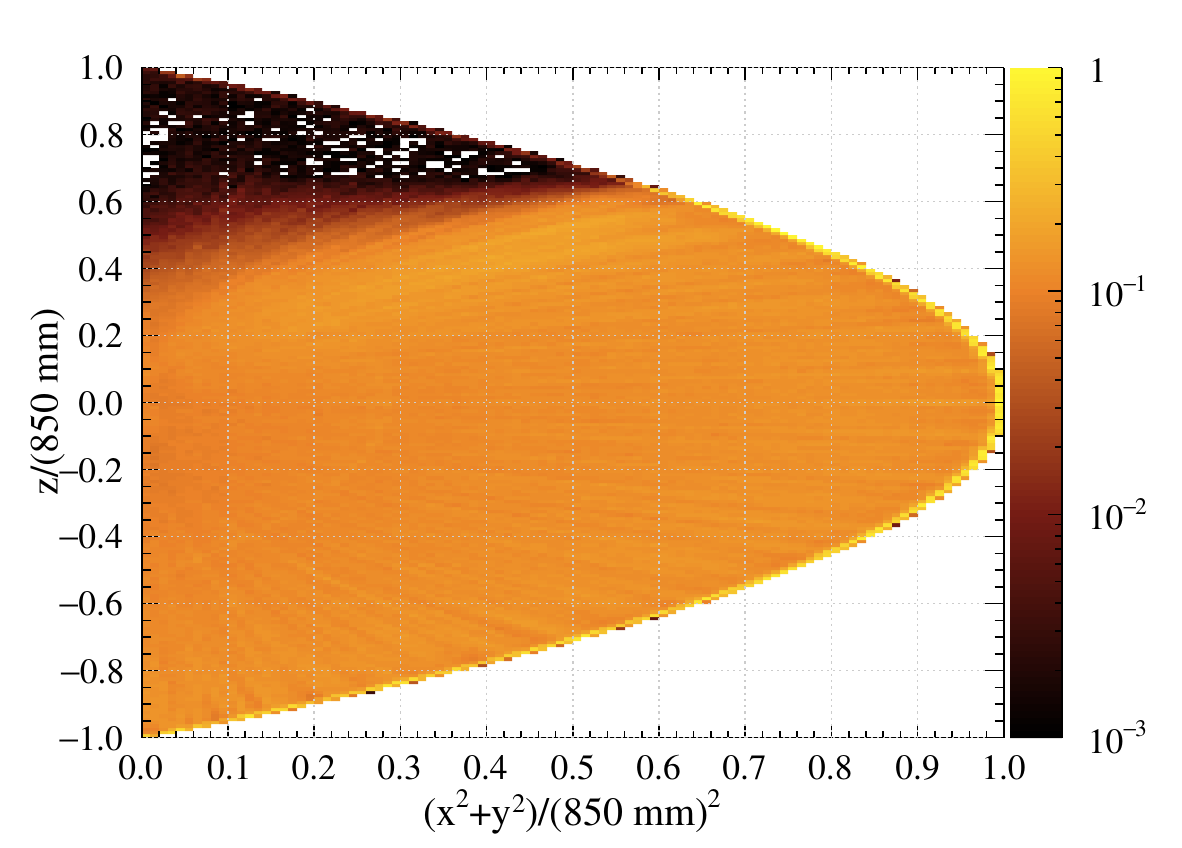}
    \includegraphics[scale=0.38]{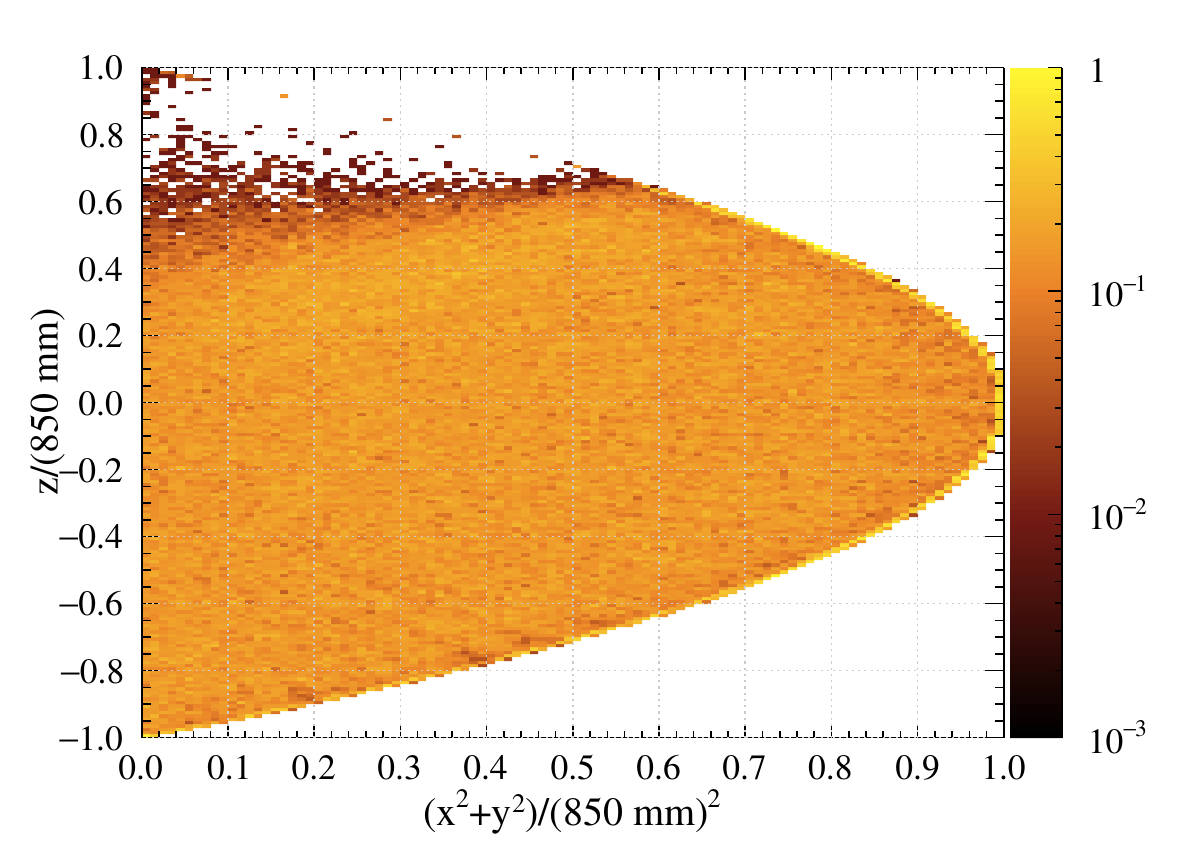}
    \caption{Reconstructed distribution shown in $z/(850$ mm) vs $(x^2+y^2)/(850\,\rm{mm})^2$ comparing $^{39}\rm{Ar}$ data (left) to $^{40}\rm{Ar}$ nuclear recoils (right) MC for the hit pattern position reconstruction. Distributions are normalized by their respective maximum values.}
    \label{fig:ZbyR0_vs_RhobyR0-square}
\end{figure}

Figure~\ref{fig:ZbyR0_vs_RhobyR0-square} shows the uniformity of reconstructed positions from the hit pattern algorithm for $^{39}\rm{Ar}$ data and $^{40}\rm{Ar}$ nuclear recoil MC data. 

\subsection{Position Reconstruction of Neck Events}
Time-of-flight and hit pattern position reconstruction algorithms use the hypothesis that an event comes from a single flash of isotropic light in LAr. Thus, the two algorithms should agree when the hypothesis is true, such as for WIMPs or $^{39}\rm{Ar}$ ER events. However, for neck events, the light is emitted above the LAr surface and shadowed by the flow-guides:
the result is a very different time-of-flight distributions and a hit pattern concentrated at the bottom of the detector.
\begin{figure}[htbp]
    \centering
    \includegraphics[scale=0.38]{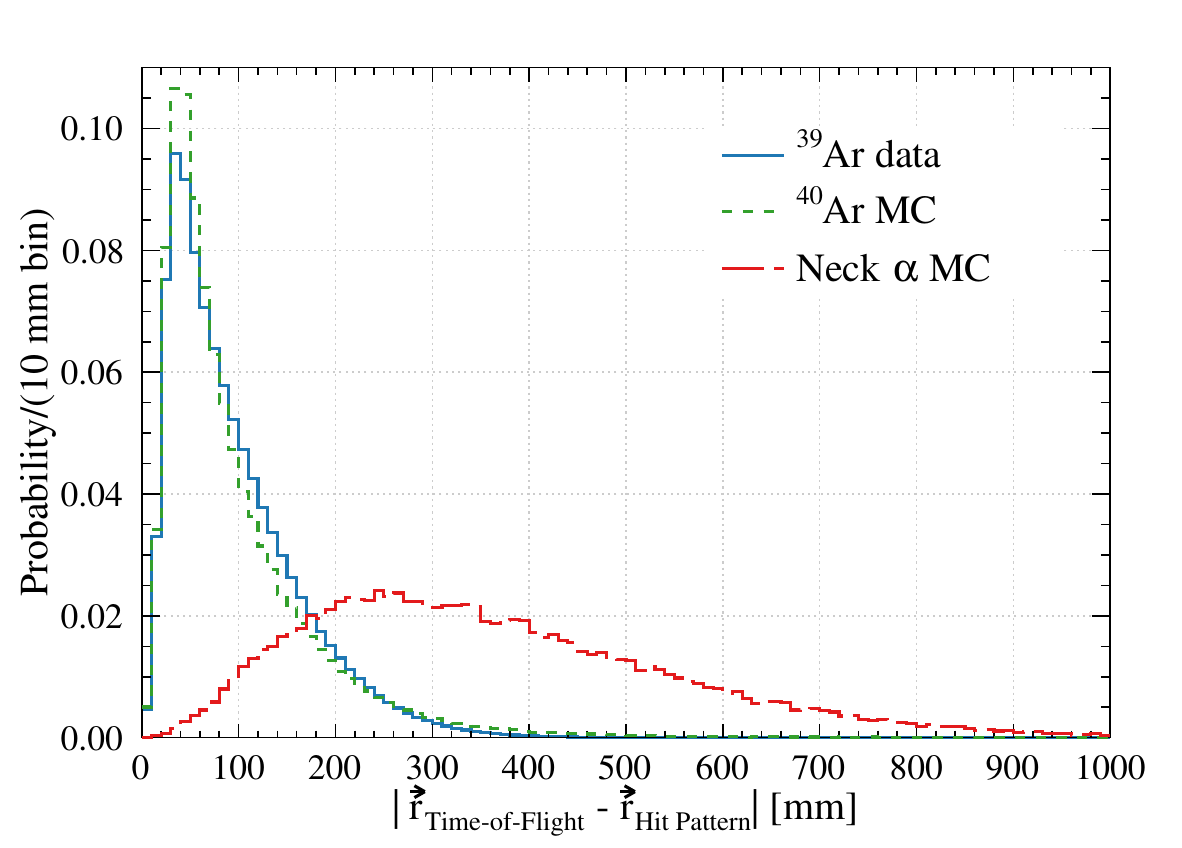}
    \includegraphics[scale=0.38]{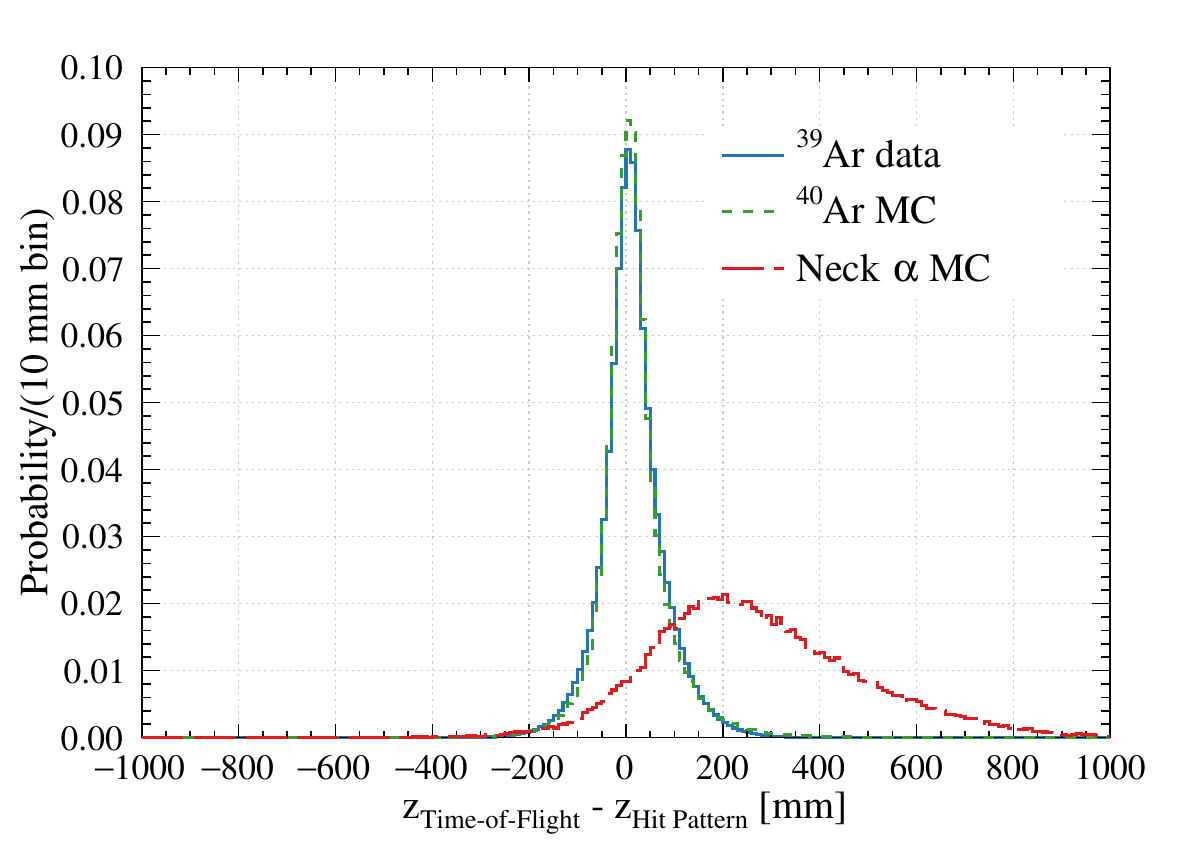}
    
    \caption{Difference between the reconstructed vertex positions (left) and vertical positions (right) between the time-based and hit-pattern based algorithms in $^{39}\rm{Ar}$ data, $^{40}\rm{Ar}$ NR simulations, and neck alpha simulations. 
    }
    \label{fig:deltaZ_and_delR_MBL_TF2}
\end{figure}
Figure~\ref{fig:deltaZ_and_delR_MBL_TF2} shows the difference between the two algorithms' reconstructed $z$ positions and vertex positions. 
$^{39}\rm{Ar}$ data, simulated WIMP-like events ($^{40}\rm{Ar}$ NR events), and simulated alpha events from the neck region are selected according to the WIMP ROI. 
It is shown that the reconstructed vertical positions agree within 35~mm for 50\% of $^{39}$Ar events in data and simulated WIMPs. For simulated neck alpha decays, the reconstructed vertical positions of the two algorithms are remarkably different, with the time-of-flight algorithm systematically reconstructing these events much closer to the top of the detector than the hit pattern algorithm. 

\begin{figure}[htbp]
    \centering
    \begin{subfigure}[b]{0.49\textwidth}
        \includegraphics[scale=0.38]{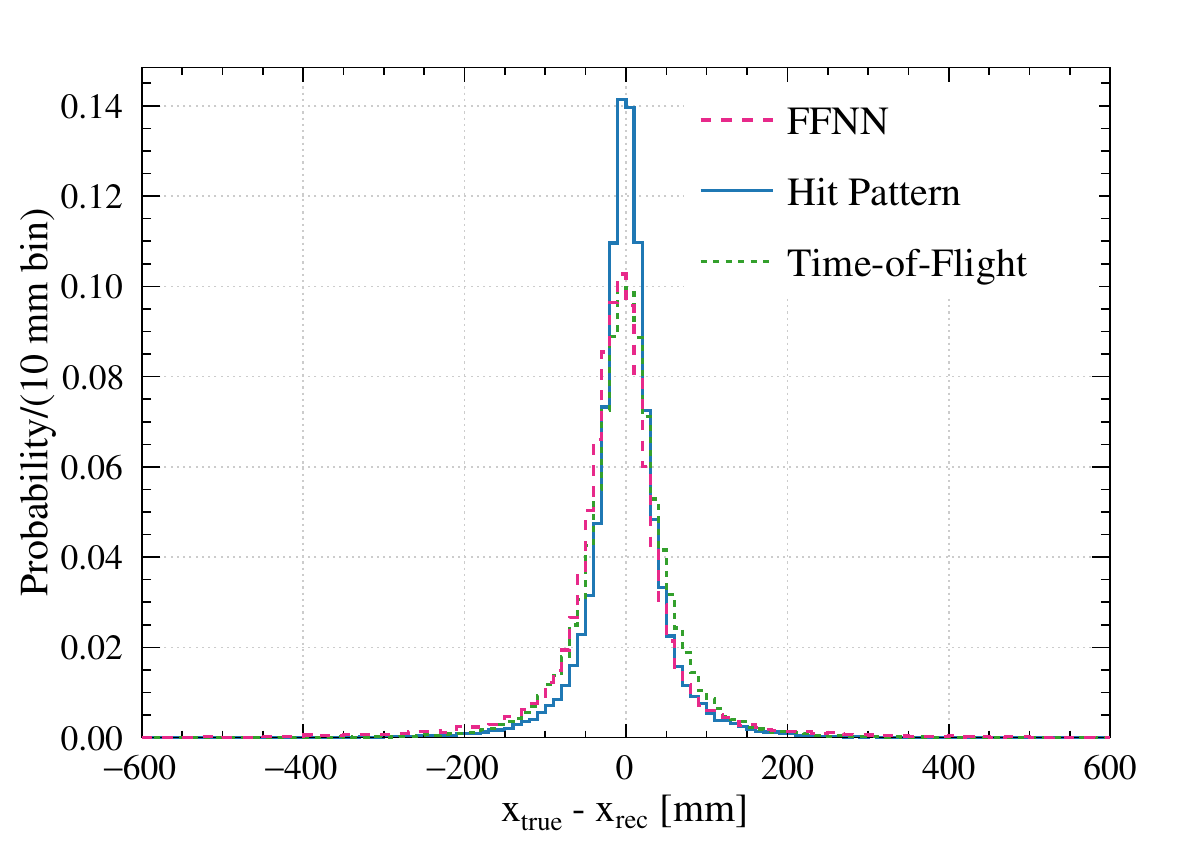}
        \caption{Error in $x$ for events originating in liquid argon.}
        \label{delx_lar}
    \end{subfigure}
    \begin{subfigure}[b]{0.49\textwidth}
        \includegraphics[scale=0.38]{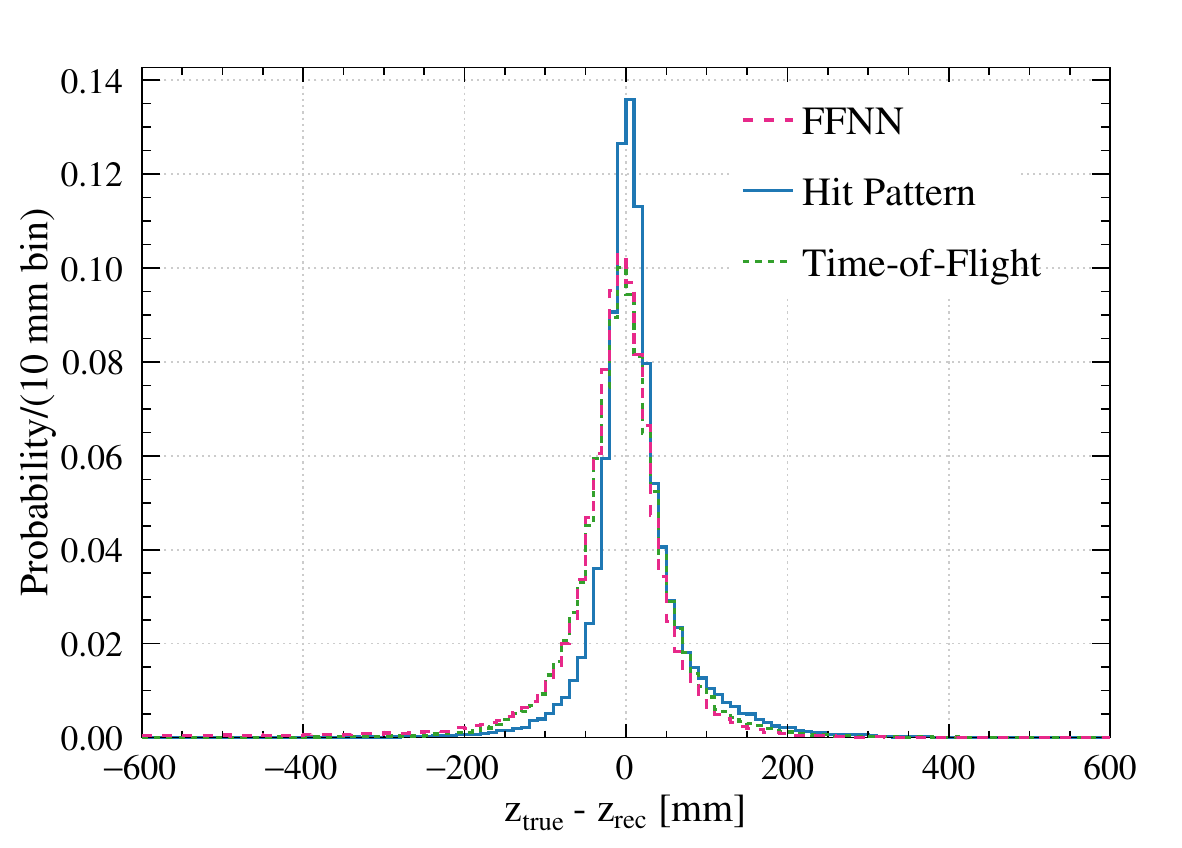}
        \caption{Error in $z$ for events originating in liquid argon.}
        \label{delz_lar}
    \end{subfigure}
    \vspace{0.5pt}
    \begin{subfigure}[b]{0.49\textwidth}
        \includegraphics[scale=0.38]{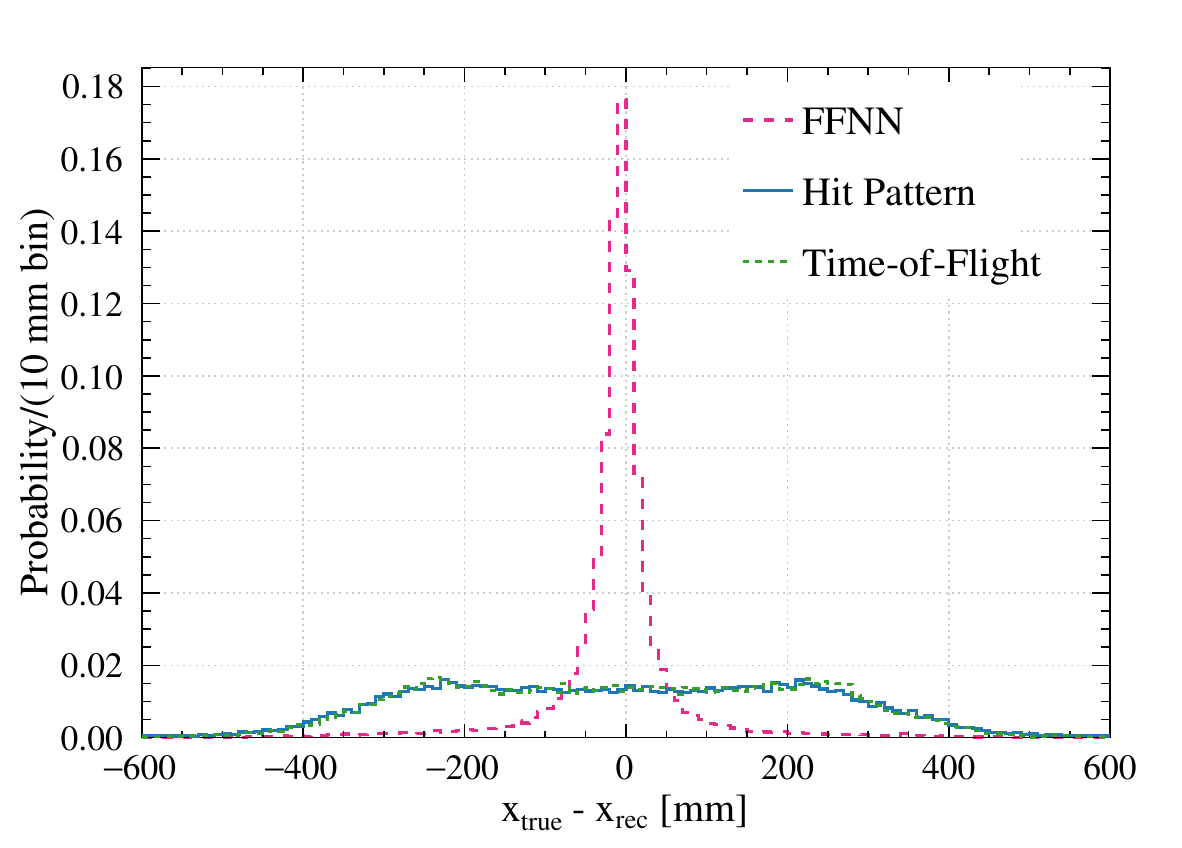}
        \caption{Error in $x$ for events originating in the neck.}
        \label{delx_neck}
    \end{subfigure}
    \begin{subfigure}[b]{0.49\textwidth}
        \includegraphics[scale=0.38]{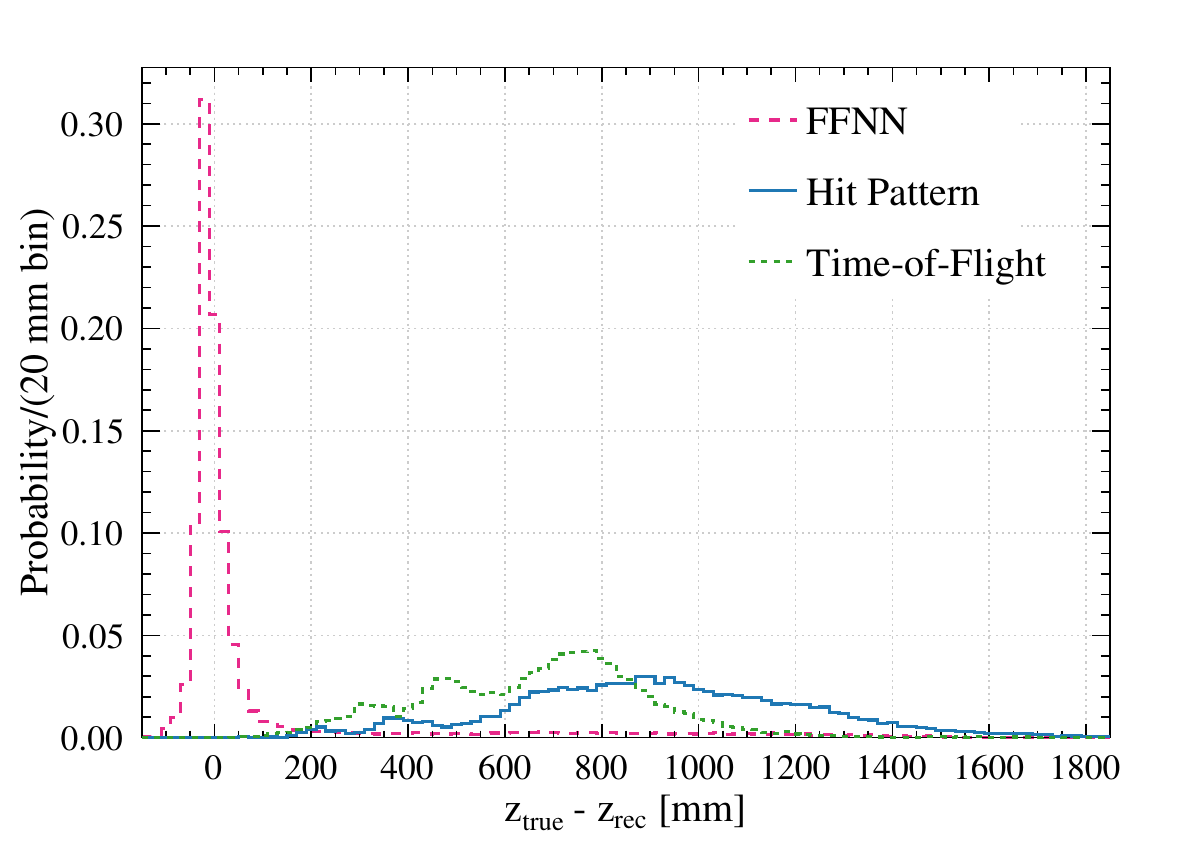}
        \caption{Error in $z$ for events originating in the neck.}
        \label{delz_neck}
    \end{subfigure}
    \caption{Distributions of the error in reconstructing $x$ ($y$ is analogous by symmetry) and $z$, 
    estimated in simulated events as the difference between reconstructed and MC "true" position.
    }
    \label{fig:delta_xz_ML_MC}
\end{figure}
%
%


In order to compare the three models in reconstructing the position of events in LAr and in the neck,
the difference between the MC ``true'' position $\vec{x}_{\rm{true}}$ and the reconstructed position $\vec{x}_{\rm{rec}}$ are calculated.
Events in LAr are defined as the region $-850 \,\rm{mm} < z_{\rm{true}} \leq 550 \,\rm{mm}$ while
the neck region is $z_{\rm{true}} > 850 \,\rm{mm}$. 
The results are presented in Figure~\ref{fig:delta_xz_ML_MC}. The new ML algorithm is comparable to the
existing hit pattern and time-of-flight algorithms in reconstructing the position in LAr, while it
outperforms the other two in the reconstruction of events in the neck.


\clearpage

\subsection{Position Reconstruction with $^{22}$Na Source Calibration Data}
\label{sec:na22}
\begin{figure} [htbp]
   \centering
   \includegraphics[width=0.8\textwidth]{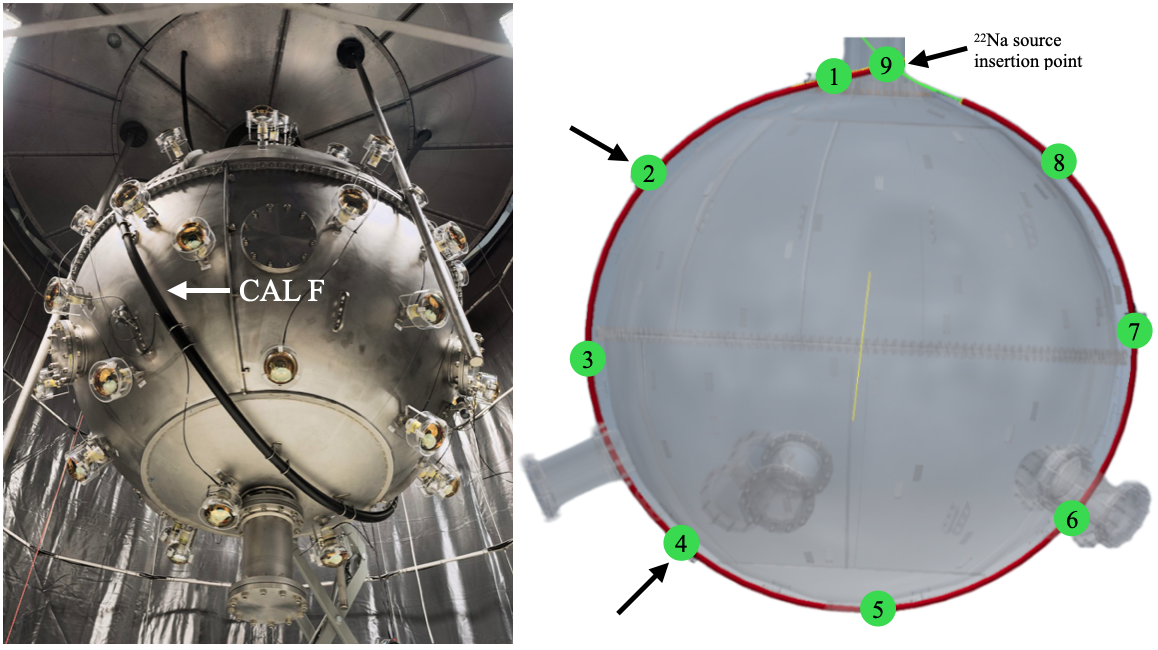}
   \caption{(Left) Picture of the steel shell containing the AV with the LAr. The CAL F pipe for the $^{22}$Na source is indicated. (Right) Drawing of the CAL F tube (in red) around the steel shell with the 9 standard positions. The two positions (2 and 4) used in this study and the insertion point are indicated with black arrows.}
   \label{fig:cal_f_outer}
\end{figure}

The possibility to employ a radioactive $\gamma$ source offers additional opportunities for testing the position reconstruction algorithms.
In particular, the source signal can be precisely tagged, effectively eliminating other background events. 
Moreover, the source can be placed above the LAr surface for investigating a region which is not covered by $^{39}$Ar decays.
The DEAP-3600 detector has a calibration polyethylene tube (known as CAL~F~\cite{DEAPdetector}) running around the steel shell holding the AV. The source is housed in a stainless steel capsule and can slide in the pipe and be placed at 9 pre-determined positions (see Figure~\ref{fig:cal_f_outer}) with an ex-situ measured uncertainty of $\sim 1$~cm.

\begin{figure} [htbp]
    \centering
    \includegraphics[scale=0.5]{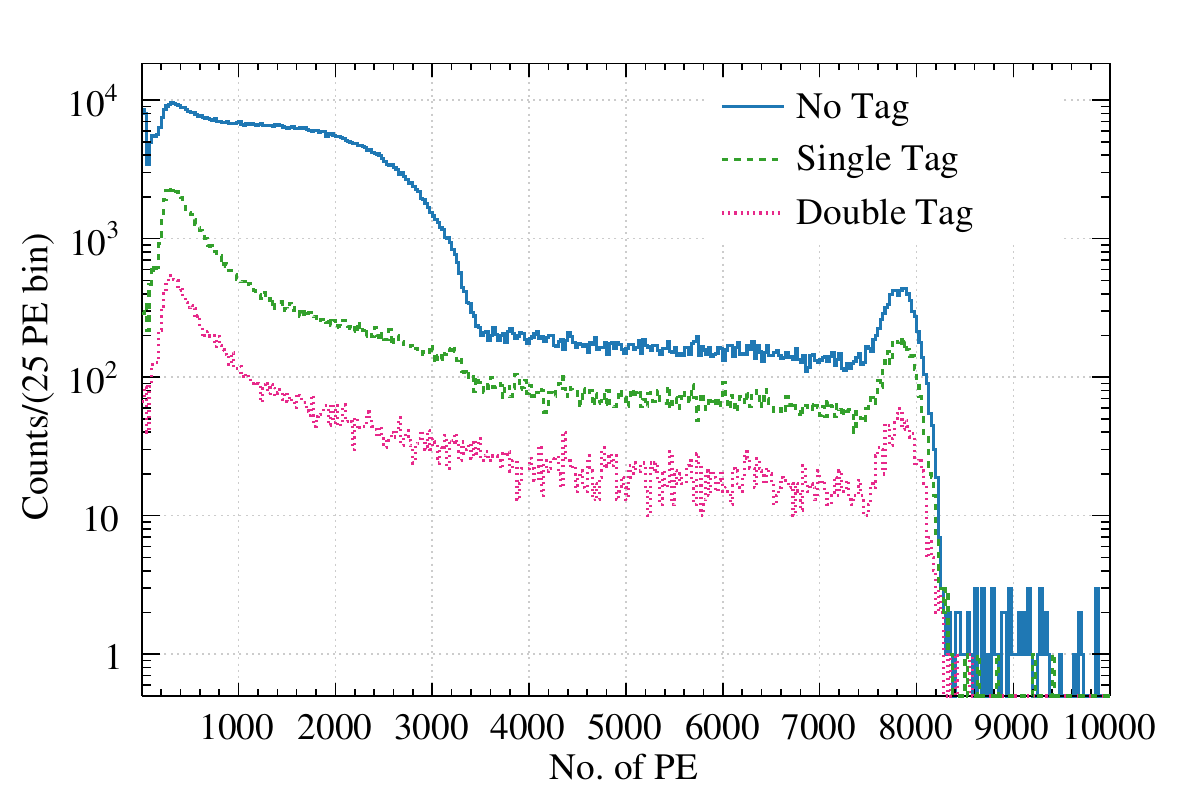}
    \caption{Energy spectrum of the $\gamma$s from the $^{22}$Na source (in PE) in LAr for different tagging configurations for position 2 (see Figure~\ref{fig:cal_f_outer}, right). }
    \label{fig:spectrum}
\end{figure}

The three $\gamma$s emitted almost simultaneously by the source allow for a tagging scheme where
two photons are detected by LYSO crystals (32 photons/keV light yield) read out by PMTs on both sides of the source, while the third one is directed towards the LAr.
The tagging algorithm relies on the energy deposited in the LYSO crystals, particularly on the timing between the event time in the LAr volume and the time when one of the tagging PMTs triggers. 
More details about this source are given in~\cite{DEAPdetector}.
The energy spectrum of the $\gamma$s from the $^{22}$Na source, interacting with the LAr volume, shows variations under three different tagging conditions (Figure~\ref{fig:spectrum}).
\begin{enumerate}
    \item No tagging by the calibration PMTs: this allows the observation of the $^{39}$Ar beta decay spectrum on top of the $^{22}$Na events; 
    \item Tagging by only one of the calibration PMTs: this occurs for instance, when the \mbox{1.27~MeV} $\gamma$ is detected by one of the calibration PMTs, one \mbox{511~keV} $\gamma$ escapes detection, and the other \mbox{511~keV} $\gamma$ reaches the LAr volume, resulting in the observed peak at \mbox{511~keV}
    ($\sim$3200~PE in Figure~\ref{fig:spectrum}, green spectrum);
    \item Double tagging: this represents the most stringent condition, where both \mbox{511~keV} $\gamma$s are tagged by the calibration PMTs, and the \mbox{1.27~MeV} interacts with the LAr ($\sim$8000~PE in Figure~\ref{fig:spectrum}).
\end{enumerate}

The full energy peak of the \mbox{1.27~MeV} $\gamma$ is visible for all the tagging conditions while the \mbox{511~keV} peak is clearly visible only for the single tag configuration.

To study the performance of the hit pattern and time-of-flight reconstruction algorithms, MC simulations were performed with the $^{22}$Na source at positions 2 and 4 (Figure~\ref{fig:cal_f_outer}). The chosen positions are below and above the LAr surface.

A total of 1.2$\times$10$^{7}$ and 4.5$\times$10$^{6}$ events were generated from the $^{22}$Na capsule in position 2 and 4, respectively. 
The difference in the number of simulated events arises from variations in the number of events that trigger the detector, which depend on their position due to the detector's geometry

\begin{figure}[htbp]
    \centering
    \begin{subfigure}[b]{0.45\textwidth}
        \includegraphics[scale=0.35]{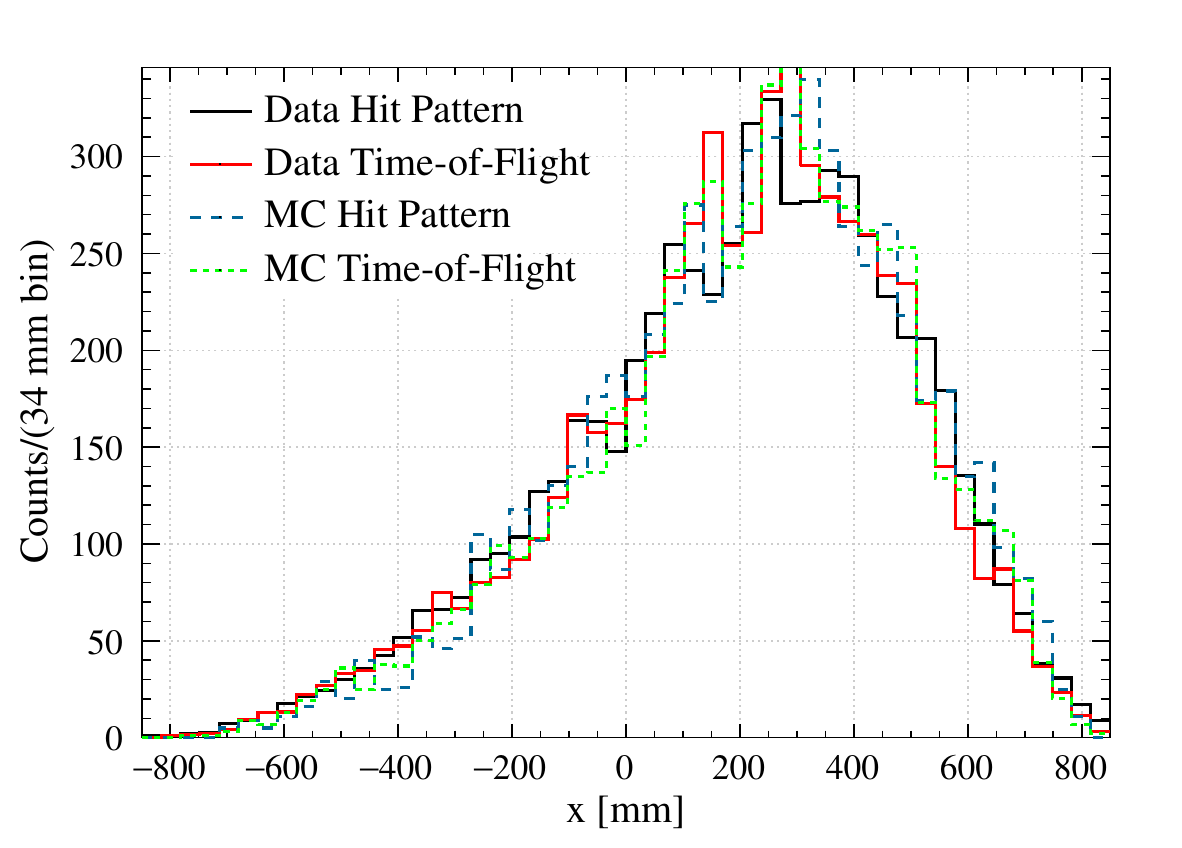}
        \caption{}
        \label{fig:pos2_x}
    \end{subfigure}
    \begin{subfigure}[b]{0.45\textwidth}
        \includegraphics[scale=0.35]{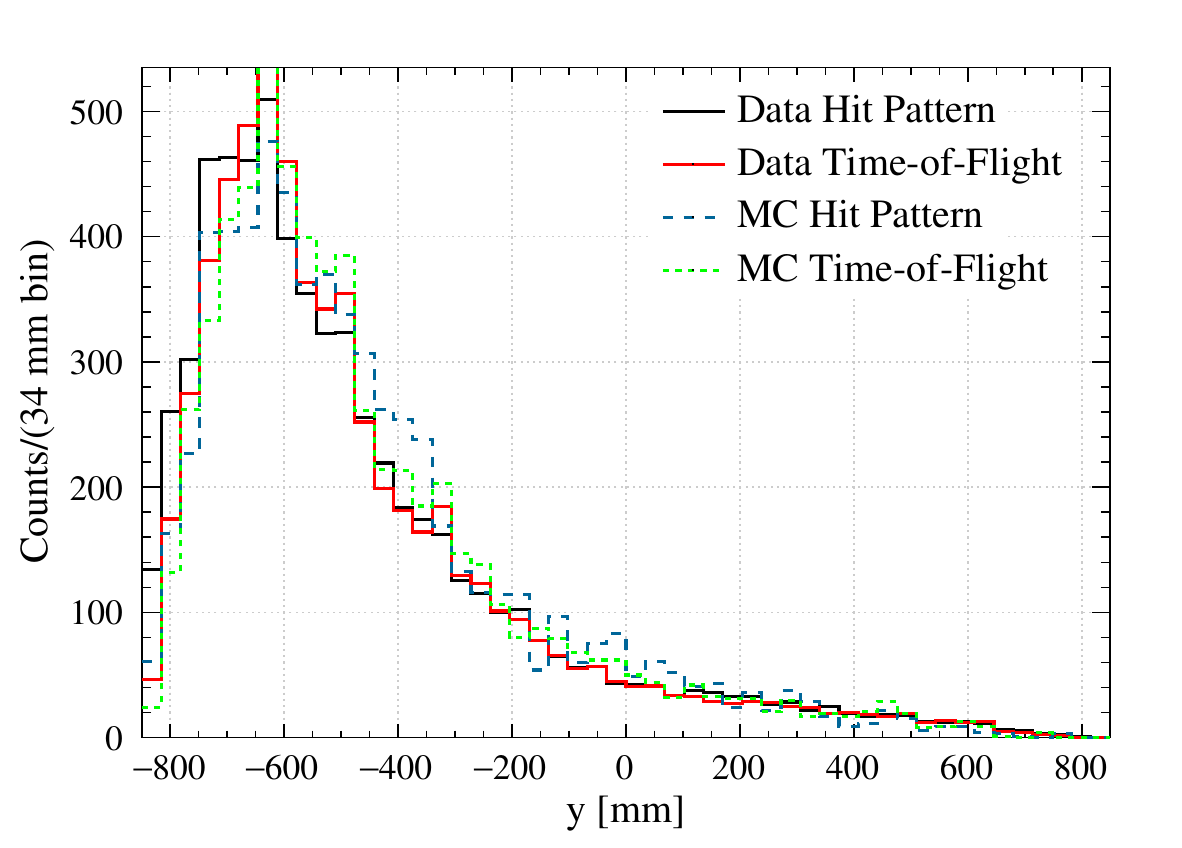}
        \caption{}
        \label{fig:pos2_y}
    \end{subfigure}
    \vspace{0.5pt} 
    \begin{subfigure}[b]{0.45\textwidth}
        \includegraphics[scale=0.35]{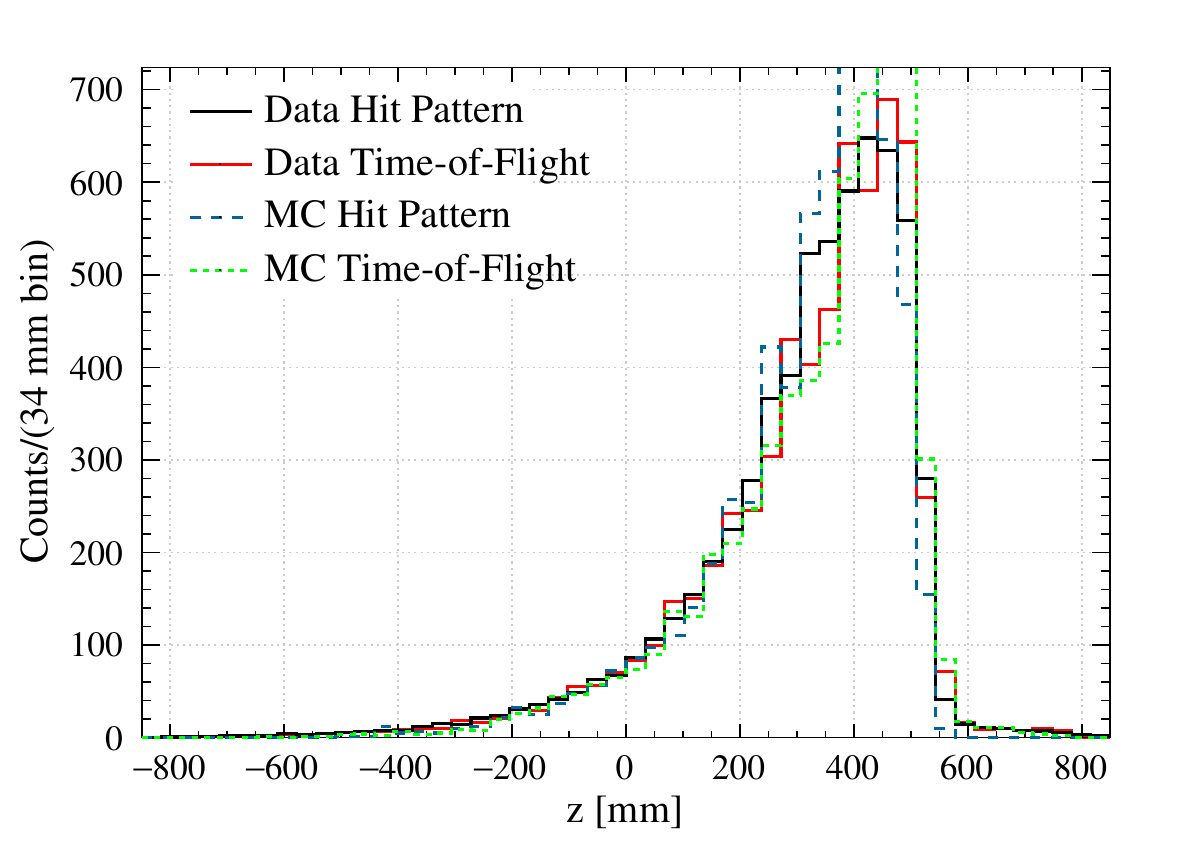}
        \caption{}
        \label{fig:pos2_z}
    \end{subfigure}
    \begin{subfigure}[b]{0.45\textwidth}
        \includegraphics[scale=0.35]{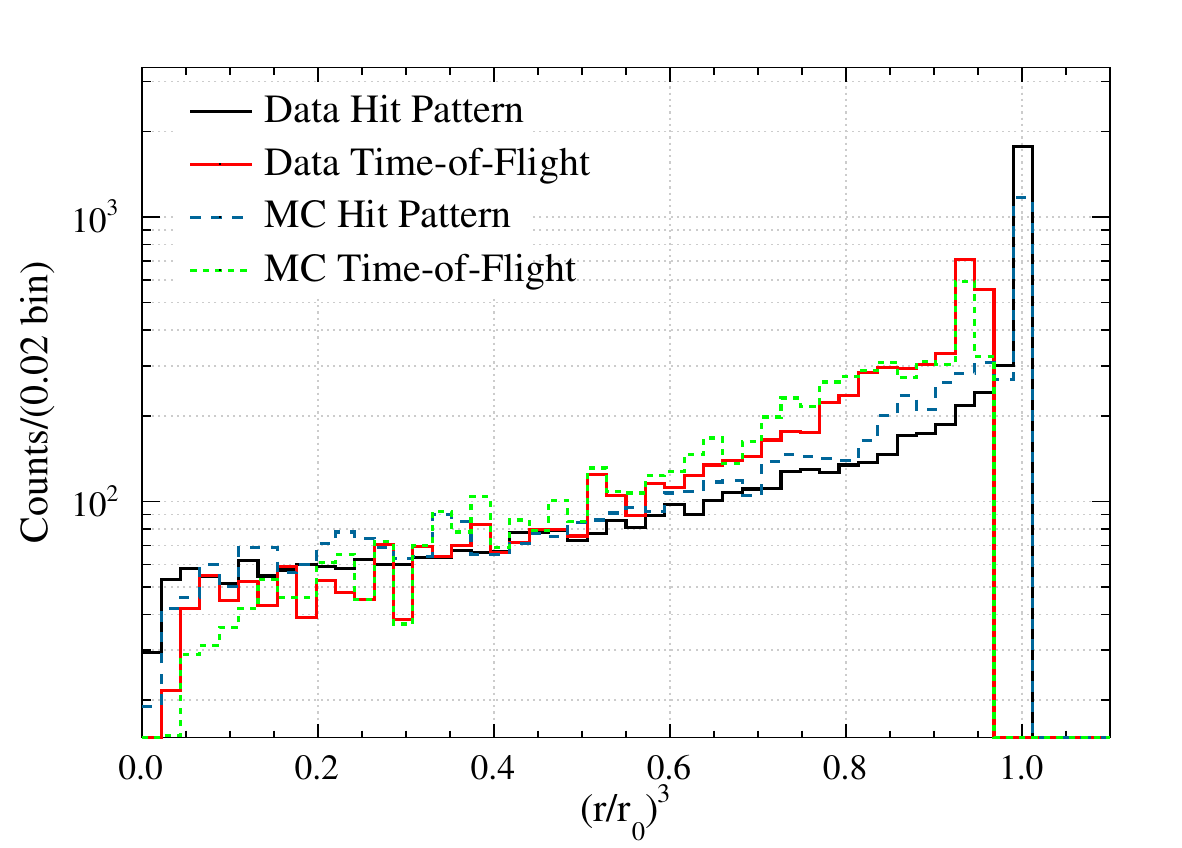}
        \caption{}
        \label{fig:pos2_r2}
    \end{subfigure}
    \caption{Reconstructed coordinates and normalized cubed radius, $(r/850\,\rm{mm})^3$ for events from position 2: data (black) and MC (dashed blue) reconstructed with the hit-pattern algorithm, data (red) and MC (dashed green) reconstructed with the time-of-flight algorithm. }
    \label{fig:pos_rec_2}
\end{figure}

\begin{figure}[htbp]
    \centering
    \begin{subfigure}[b]{0.45\textwidth}
        \includegraphics[scale=0.35]{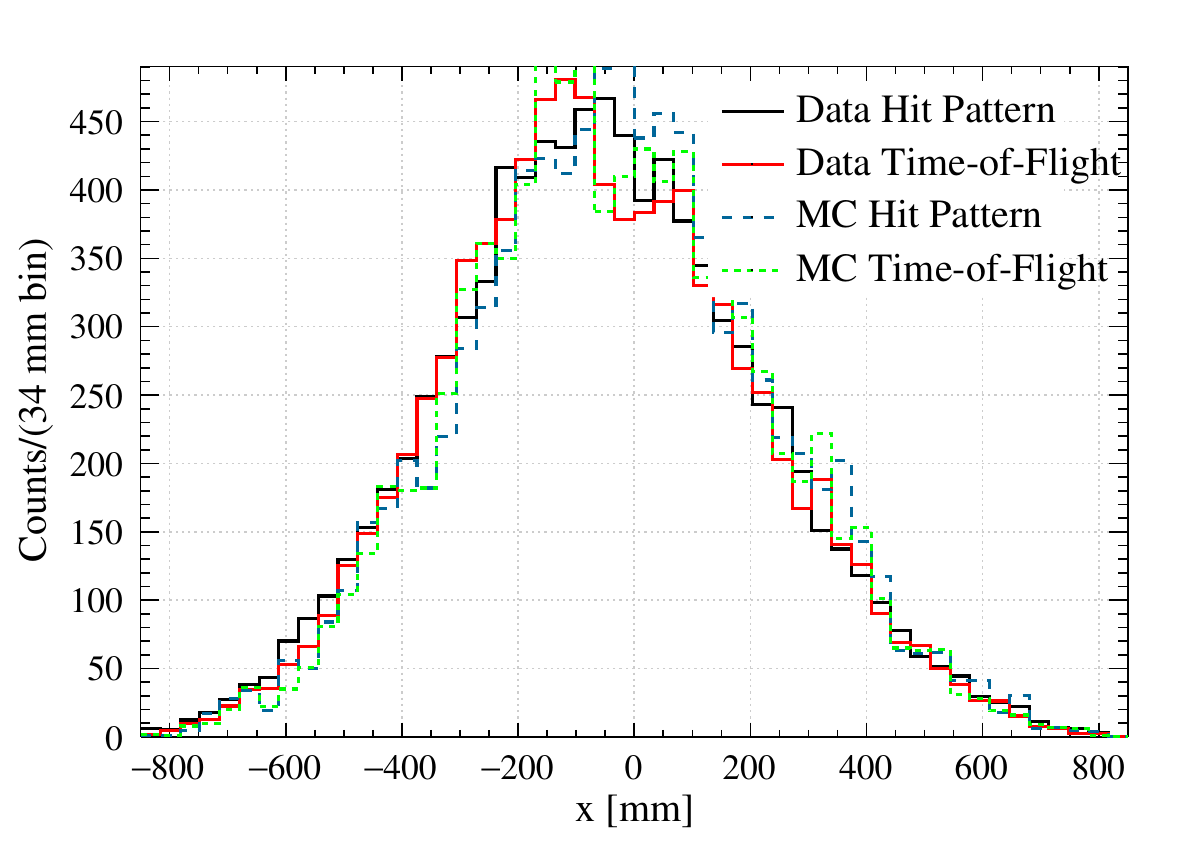}
        \caption{}
        \label{fig:pos4_x}
    \end{subfigure}
    \begin{subfigure}[b]{0.45\textwidth}
        \includegraphics[scale=0.35]{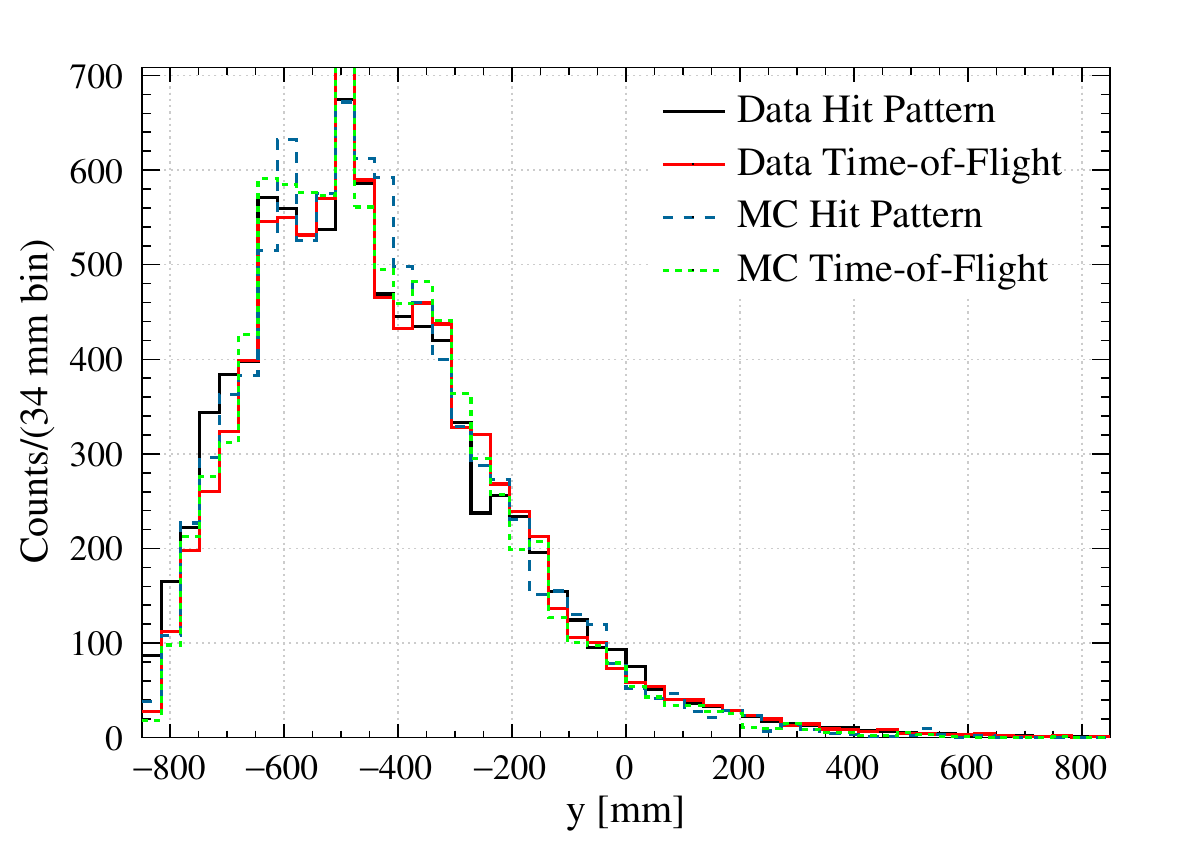}
        \caption{}
        \label{fig:pos4_y}
    \end{subfigure}
    \vspace{0.5 pt}
    \begin{subfigure}[b]{0.45\textwidth}
        \includegraphics[scale=0.35]{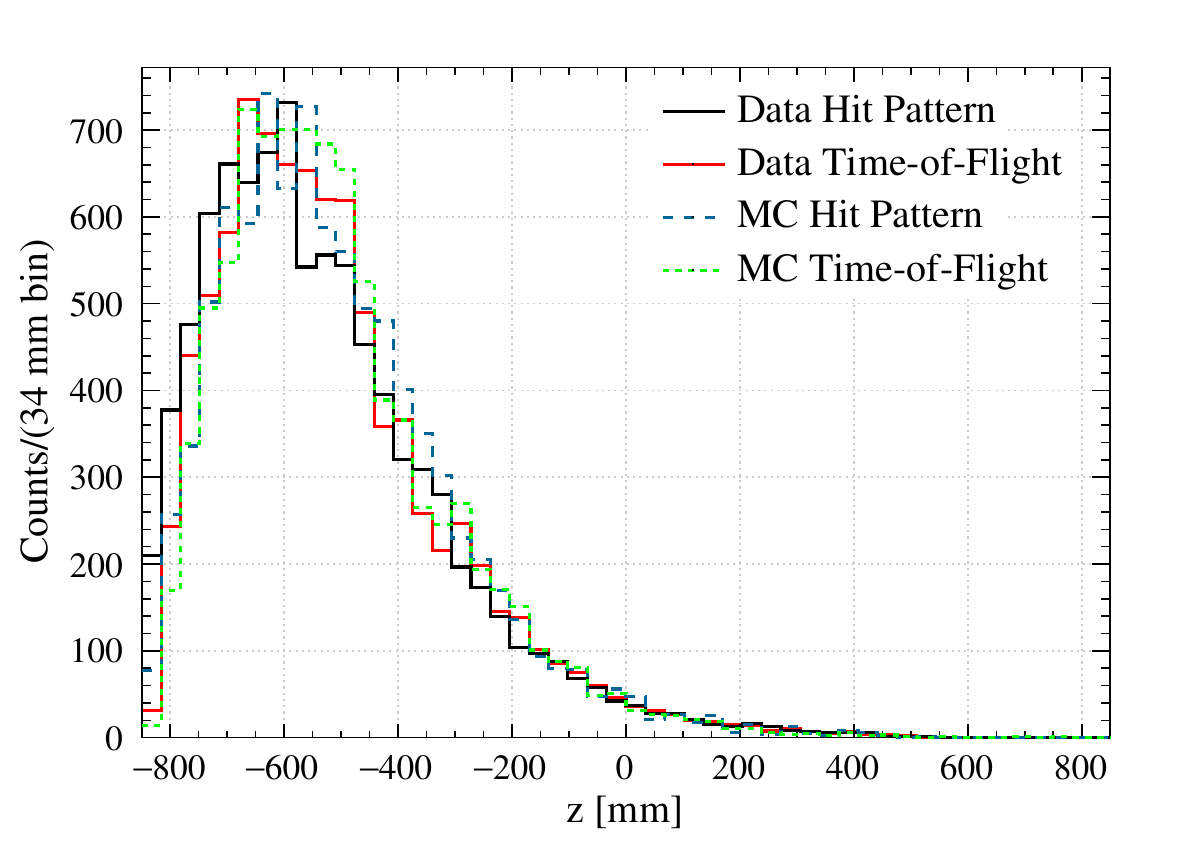}
        \caption{}
        \label{fig:pos4_z}
    \end{subfigure}
    \begin{subfigure}[b]{0.45\textwidth}    
        \includegraphics[scale=0.35]{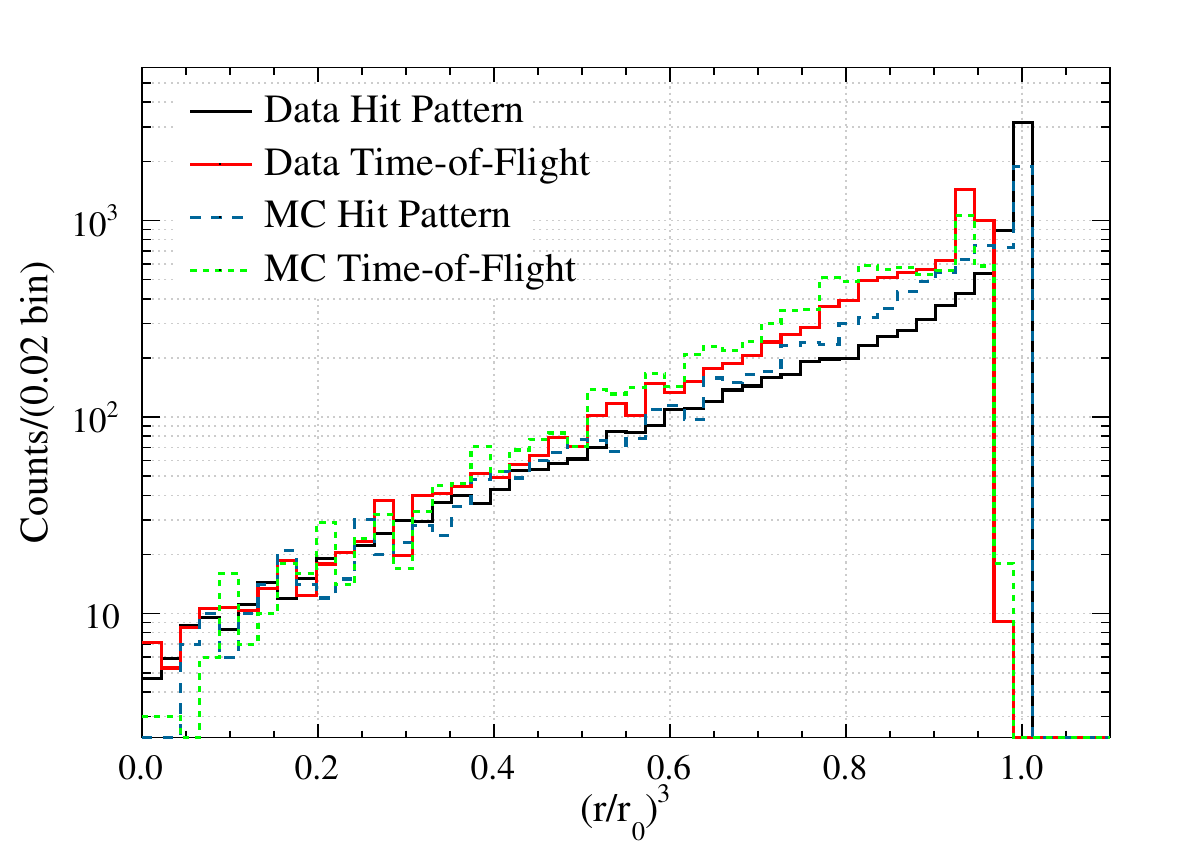}
        \caption{}
        \label{fig:pos4_r2}
    \end{subfigure}
    \caption{Reconstructed coordinates and normalized cubed radius, $(r/850\,\rm{mm})^3$ for events from position 4: data (black) and MC (dashed blue) reconstructed with the hit-pattern algorithm, data (red) and MC (dashed green) reconstructed with the time-of-flight algorithm.}
    \label{fig:pos_rec_4}
\end{figure}

Reconstructed coordinates for positions 2 and 4 are shown in Figures~\ref{fig:pos_rec_2} and \ref{fig:pos_rec_4}, for data and MC, both reconstructed with the hit-pattern and time-of-flight algorithms (normalizing to the same number of events). 

A general good agreement between data and MC is observed for all coordinates but differences between
the two algorithms can be noticed. From Figures~\ref{fig:pos_rec_2}(d) and \ref{fig:pos_rec_4}(d), it is clear
that the time-of-flight algorithm tends to reconstruct events closer to the center of the detector with respect to
the hit pattern-based algorithm. 

\clearpage

\section{Position Resolution}
\label{sec:resolution}

A data-driven method, first introduced in~\cite{DEAPdm2019}, is used to quantify the position resolution obtained from the three algorithms.
According to this method, starting from a sample of $^{39}$Ar decays, we create two ``pseudo-events'' each with about half the original event's observed PE originating from the same position in the detector.  
This is done by cycling through all the PMTs that saw light in the event, then through each PMT pulse, and through each PE within pulses.  
Each such PE unit from the original event can be assigned to the first pseudo-event with 50\% probability and independently can be assigned to the second pseudo-event with 50\% probability to minimize correlations between the two pseudo-events\footnote{In general, a number $n$ of pseudo-events can be considered, but we consider $n=2$ for this analysis.}.
In this way, a PE unit from a pulse can be added to both pseudo-events, one or the other, or neither. On average, each pseudo-event has approximately half the energy of the original event and a similar spatial and time profile.

The original event and pseudo-events can then be processed by the position reconstruction algorithms.
Comparisons in reconstructed position can be drawn between the pseudo-events and the original event, since the origin of the scintillation photons in the detector is the same. 

For a large sample of events distributed throughout the detector volume and in a range of energies, the distributions of the differences in reconstructed position $(\Delta x, \Delta y, \Delta z)$ between pairs of pseudo-events provide a data-driven estimate of the position resolution.  
The resolution estimator $\overline{\delta r}$ on the event position from the origin of the detector is calculated from the resolution on measured Cartesian coordinates $(\sigma_{\rm{eff}}^x, \sigma_{\rm{eff}}^y, \sigma_{\rm{eff}}^z)$ as:
\begin{equation}
    \overline{\delta r} = \sqrt{\frac{ (\sigma_{\rm{eff}}^x)^2 + (\sigma_{\rm{eff}}^y)^2 + (\sigma_{\rm{eff}}^z)^2 }{3}}\quad,
    \label{eq:reso_radius}
\end{equation}
where the Cartesian resolution estimators are obtained from the characteristic widths of the distributions of $\Delta x$, $\Delta y$ and $\Delta z$ (see Appendix~\ref{app:eff_sigma}).

\begin{figure}[htbp]
    \centering
    \begin{subfigure}[b]{0.48\textwidth} 
        \centering
         \includegraphics[scale=0.35]{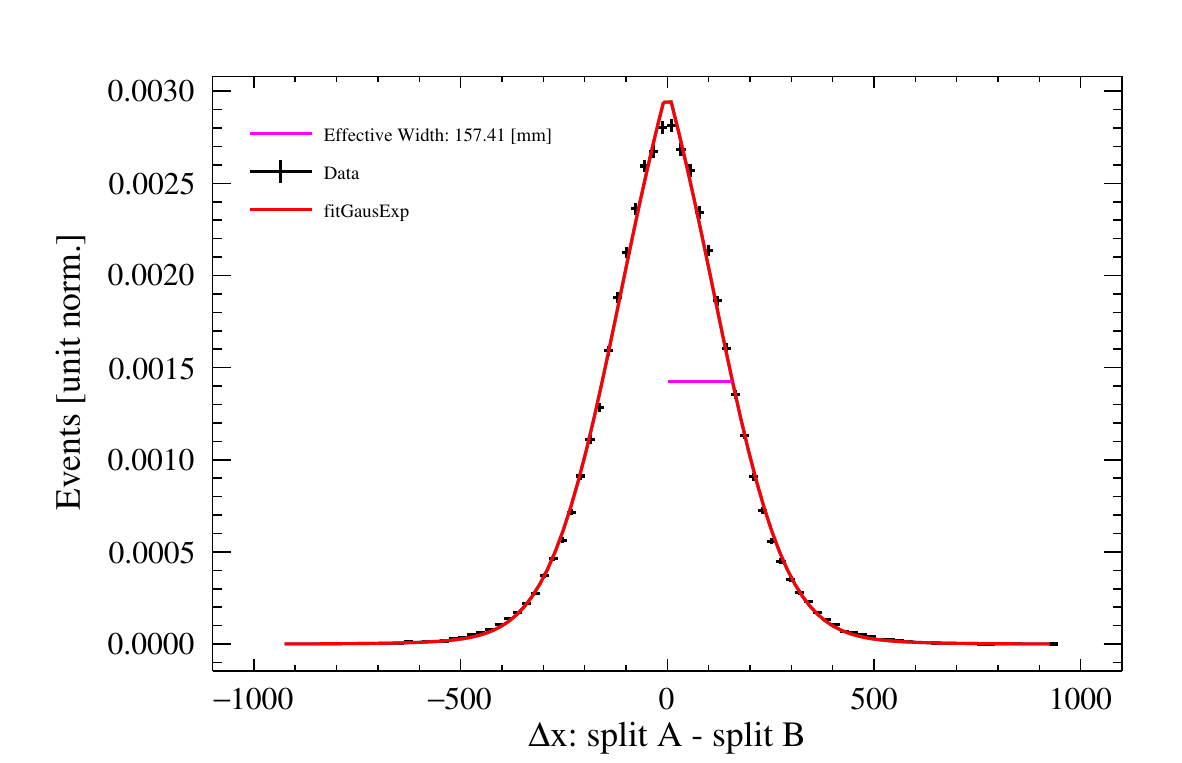}
        \caption{Hit pattern reconstructed $x$, $0<r<150$~mm.}
        \label{fig:split_event_dist_mbl_lowR}
    \end{subfigure}
    ~
    \begin{subfigure}[b]{0.48\textwidth}
        \centering
        \includegraphics[scale=0.35]{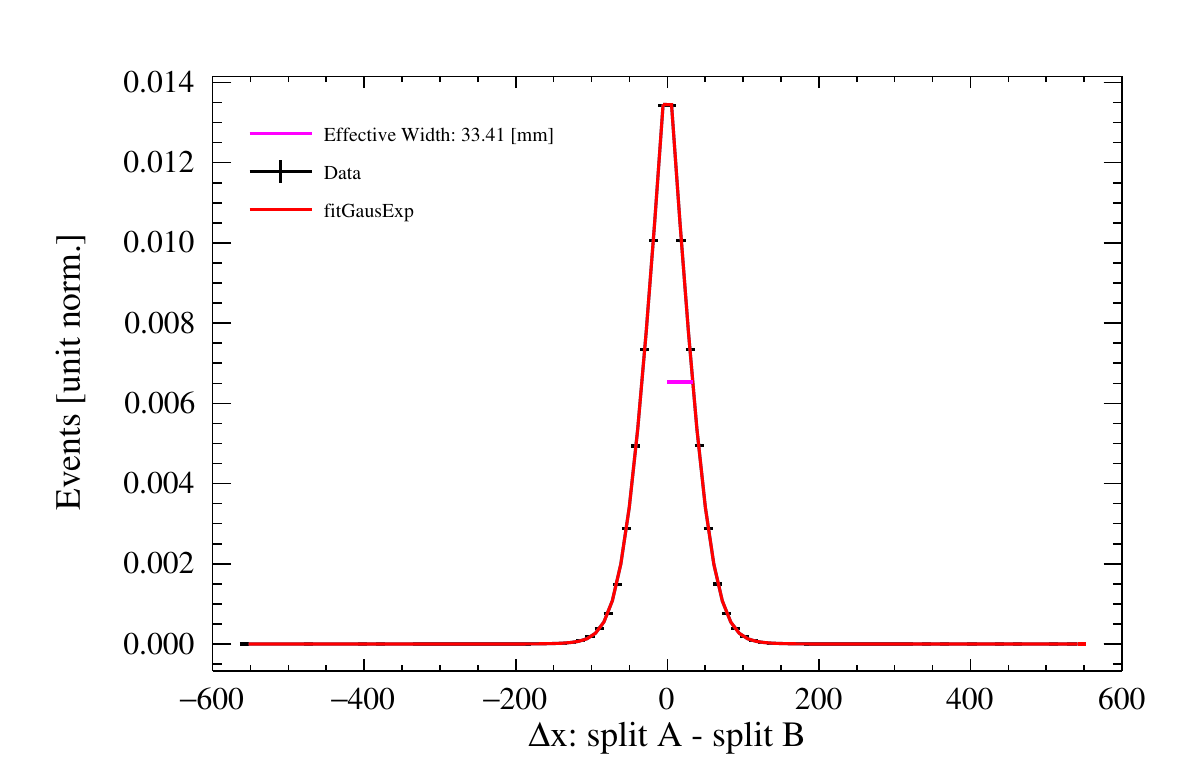}
        \caption{Hit pattern reconstructed $x$, $750<r<800$~mm.}
        \label{fig:split_event_dist_mbl_highR}
    \end{subfigure}
    ~
    \begin{subfigure}[b]{0.48\textwidth}
        \centering
        \includegraphics[scale=0.35]{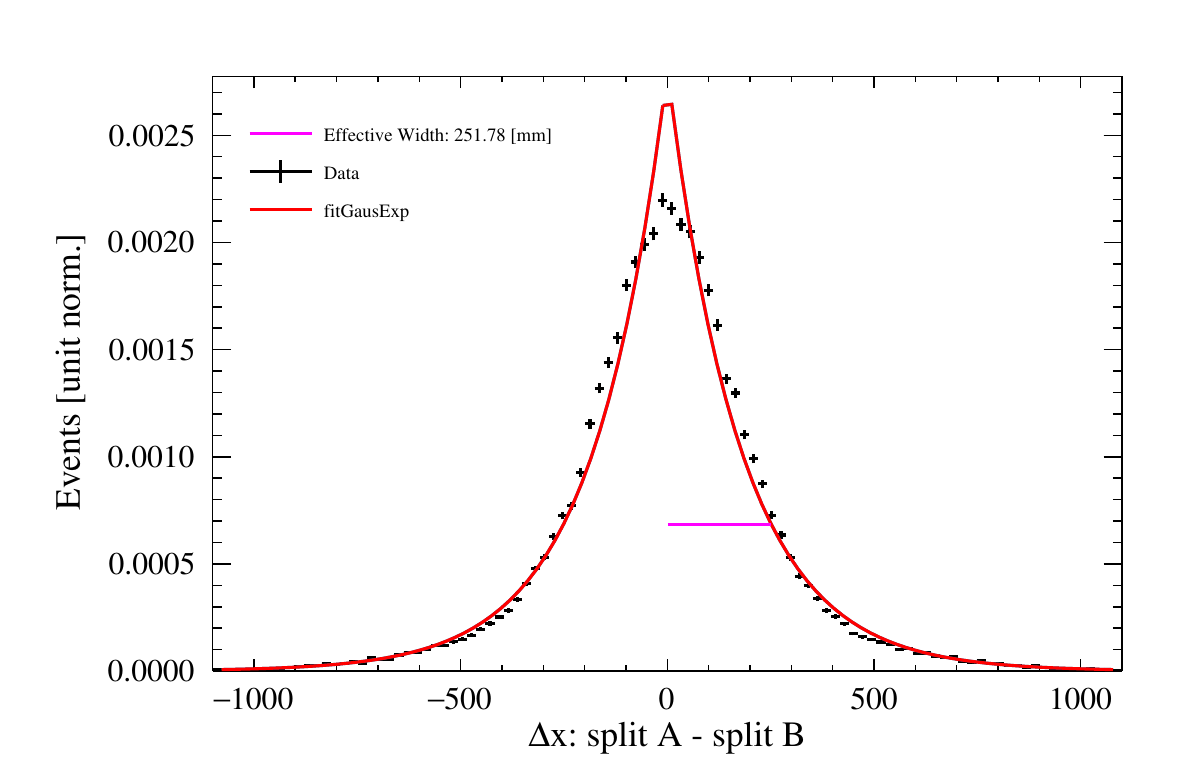}    
        \caption{Time-based reconstructed x, $0<r<150$~mm.}
        \label{fig:split_event_dist_tf2_lowR}
    \end{subfigure}
    ~
    \begin{subfigure}[b]{0.48\textwidth}
        \centering
        \includegraphics[scale=0.35]{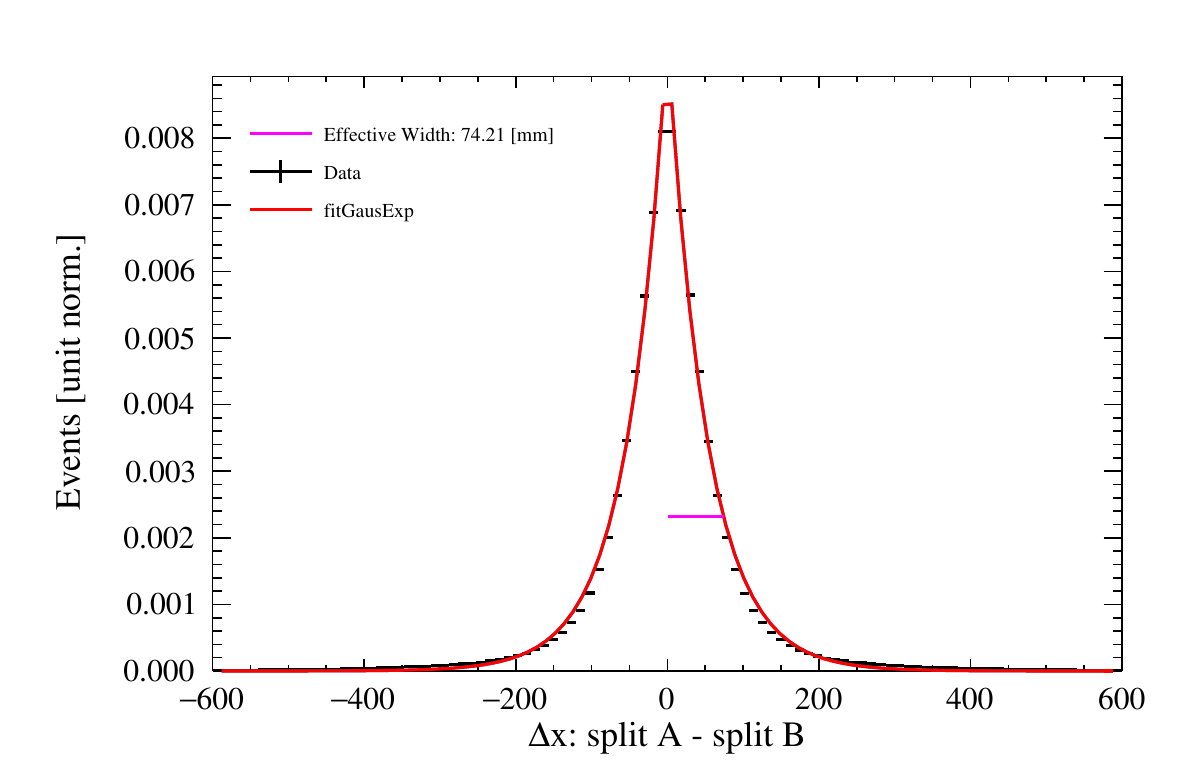}    
        \caption{Time-based reconstructed x, $750<r<800$~mm.}
        \label{fig:split_event_dist_tf2_highR}
    \end{subfigure}
    ~
        \begin{subfigure}[b]{0.48\textwidth}
        \centering
        \includegraphics[scale=0.35]{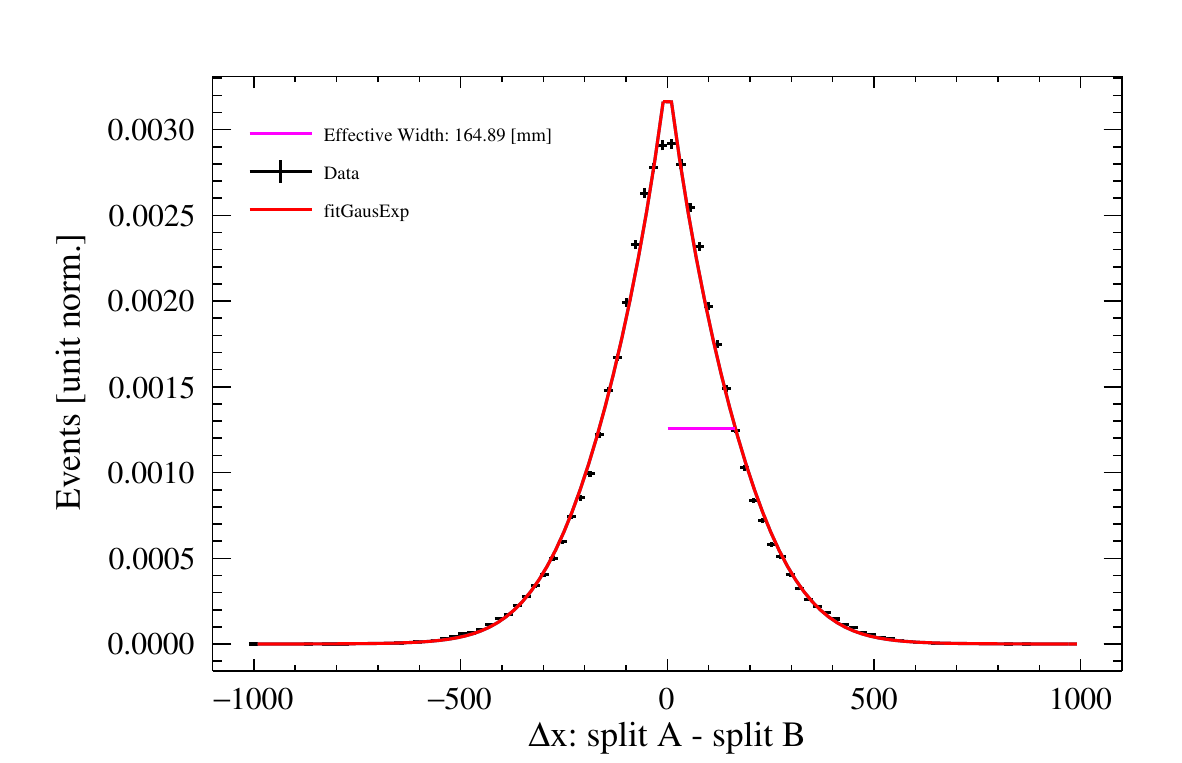}    
        \caption{Neural network reconstructed $x$, $0<r<150$~mm.}
        \label{fig:split_event_dist_nnX_lowR}
    \end{subfigure}
    ~
    \begin{subfigure}[b]{0.48\textwidth}
        \centering
        \includegraphics[scale=0.35]{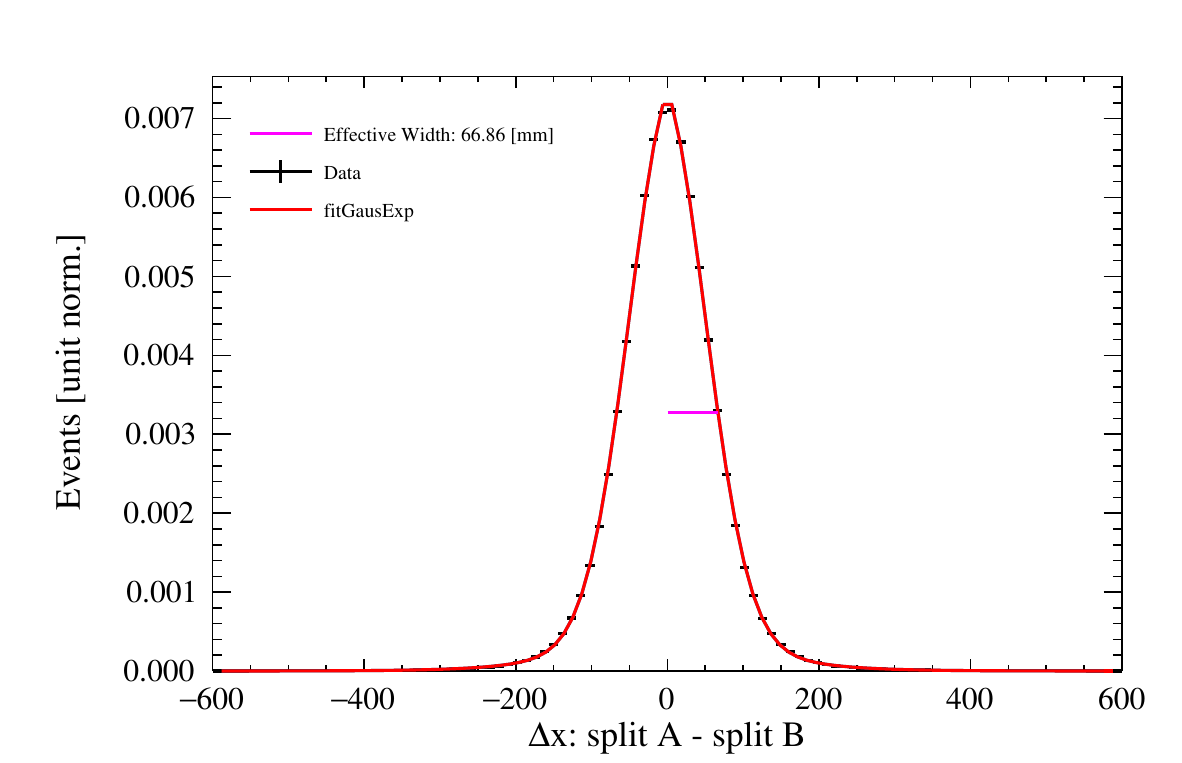}    
        \caption{Neural network reconstructed $x$, $750<r<800$~mm.}
        \label{fig:split_event_dist_nnX_highR}
    \end{subfigure}
    ~
        \begin{subfigure}[b]{0.48\textwidth}
        \centering
        \includegraphics[scale=0.35]{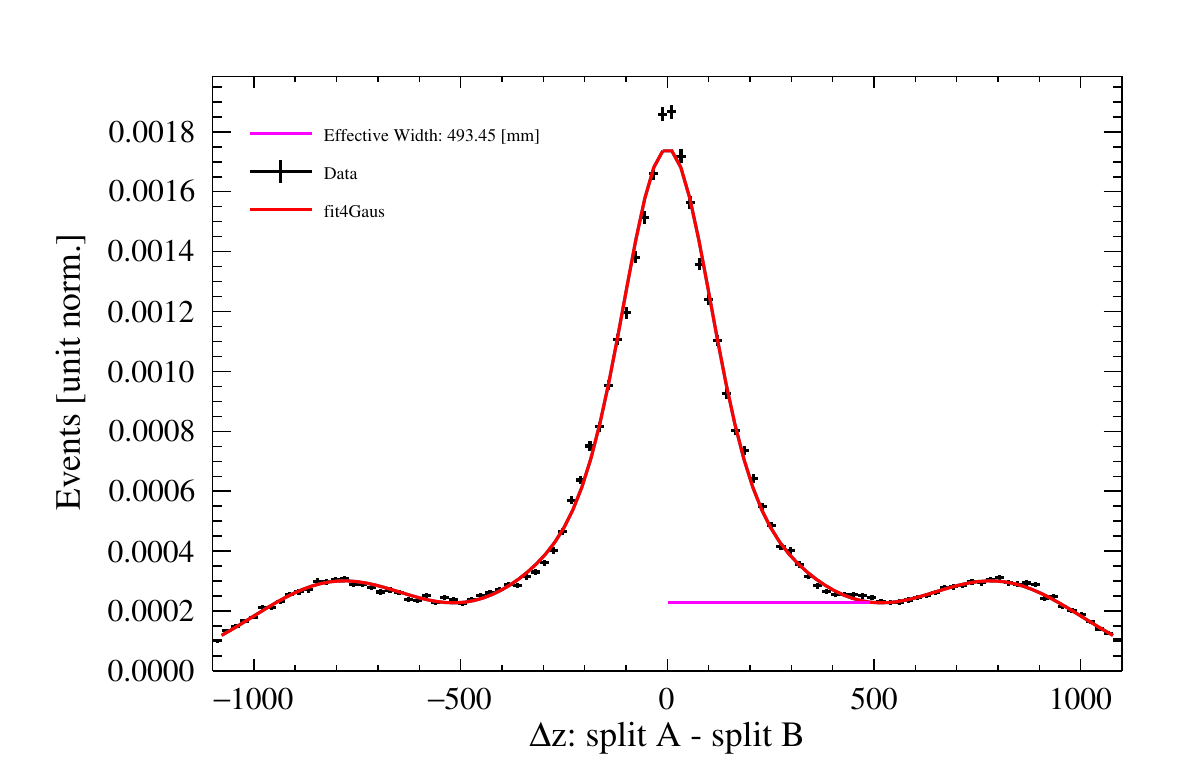}    
        \caption{Neural network reconstructed $z$, $0<r<150$~mm.}
        \label{fig:split_event_dist_NNZ_lowR}
    \end{subfigure}
    ~
    \begin{subfigure}[b]{0.48\textwidth}
        \centering
        \includegraphics[scale=0.35]{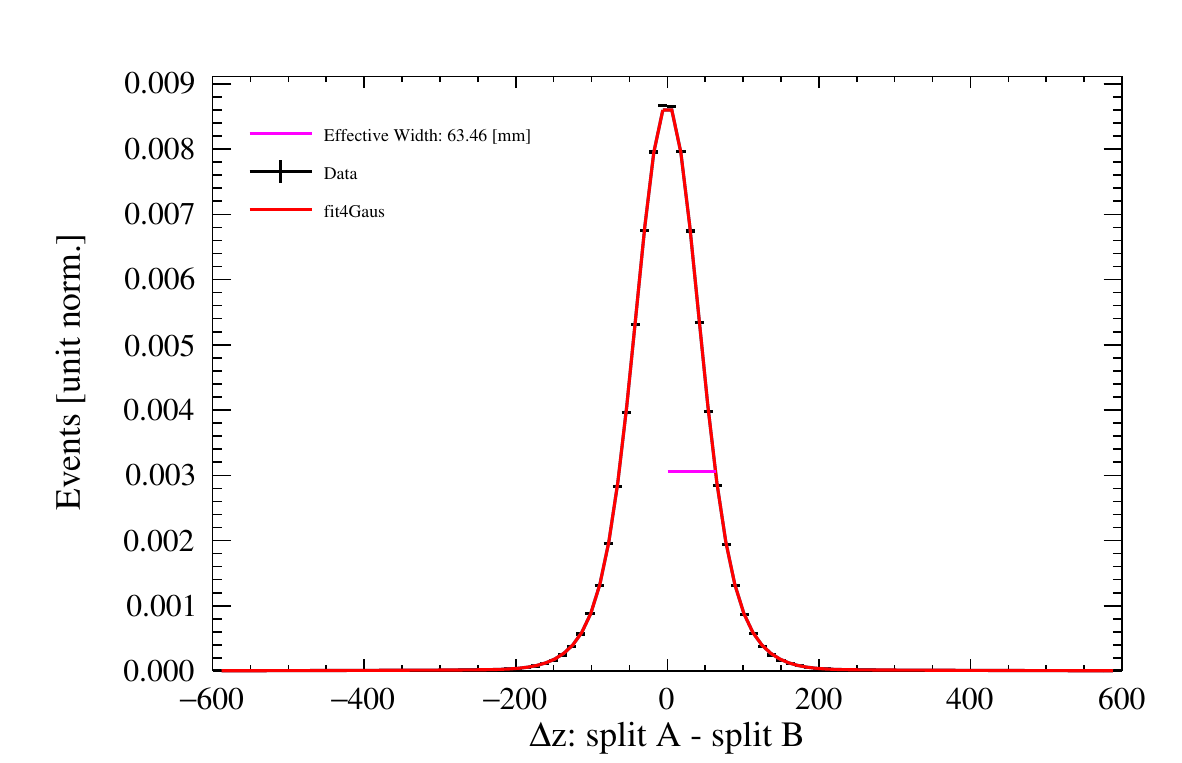}    
        \caption{Neural network reconstructed $z$, $750<r<800$~mm.}
        \label{fig:split_event_dist_nnZ_highR}
    \end{subfigure}
    ~
    \caption{Examples of the distributions of the differences in reconstructed position between pseudo-events in the 100--200 PE range, for the three algorithms. Distributions for two regions, based on the reconstructed radius of the original $^{39}$Ar decay event, are shown: near the center of the detector (left), and near the edges (right).
    }
    \label{fig:resolution_delta_distributions}
\end{figure}

\begin{figure}[h!tb]
    \centering
    \begin{subfigure}[b]{0.49\textwidth}
        \centering
        \includegraphics[scale=0.4]{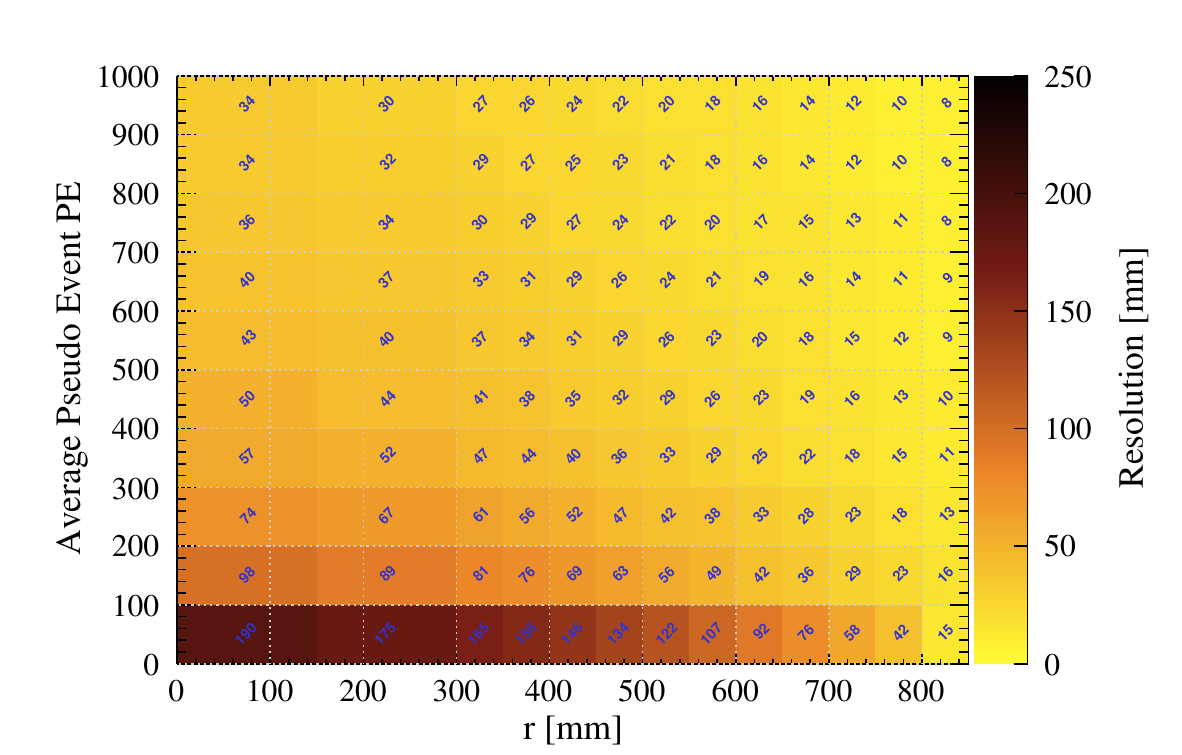}
        \caption{Likelihood hit pattern-based position reconstruction}
        \label{fig:reso_MBL}
    \end{subfigure}
    ~
    \begin{subfigure}[b]{0.49\textwidth}
        \centering
        \includegraphics[scale=0.4]{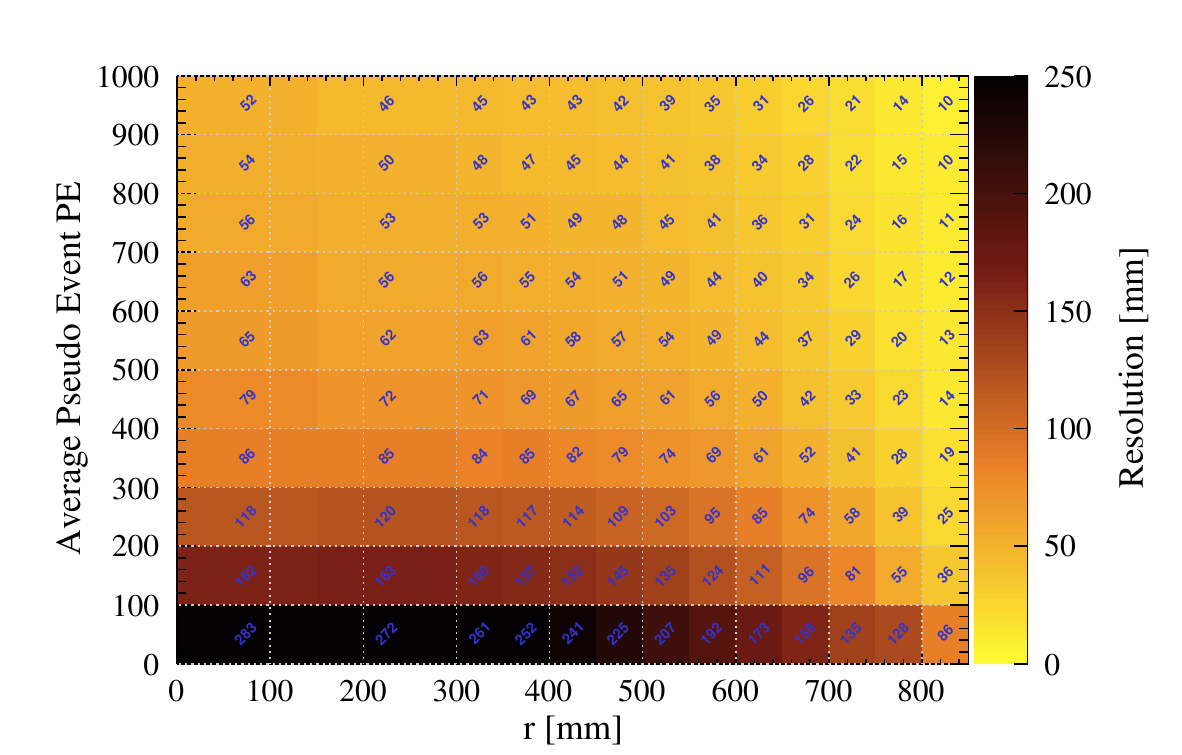}
        \caption{Likelihood time-based position reconstruction}
        \label{fig:reso_TF2}        
    \end{subfigure}
    \vspace{0.5pt}
    \begin{subfigure}[b]{0.49\textwidth}
        \centering
        \includegraphics[scale=0.4]{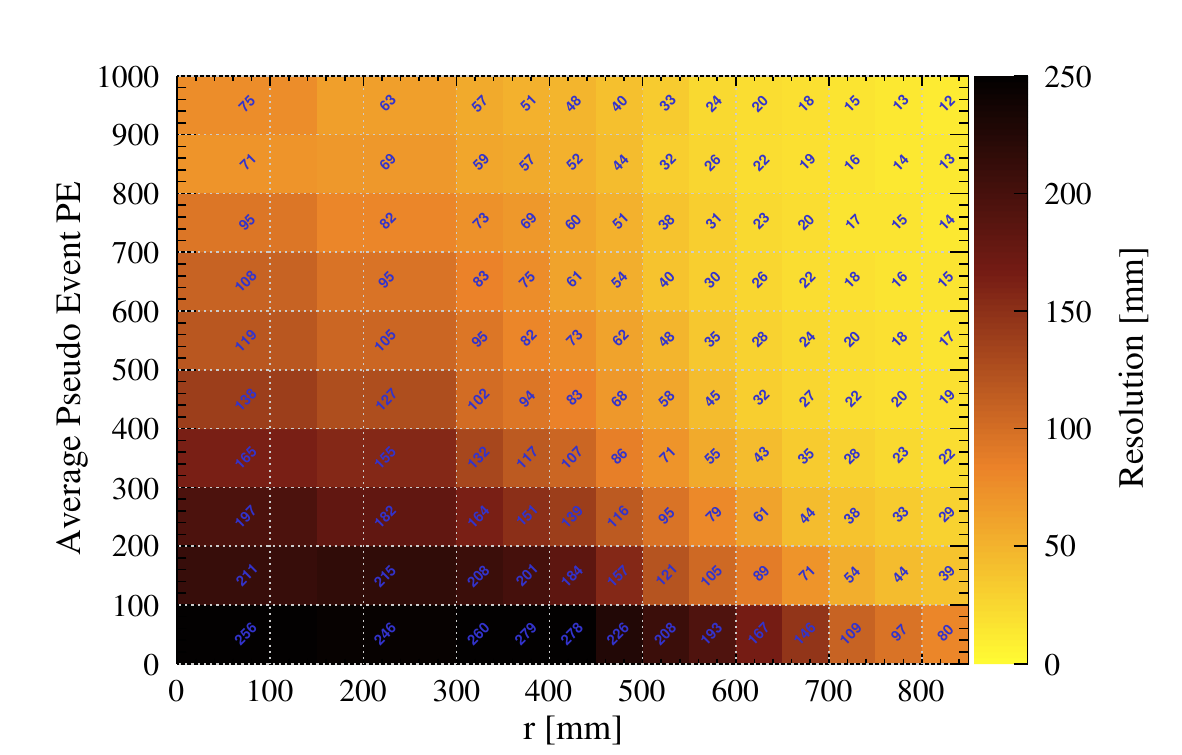}
        \caption{Neural network PE-based position reconstruction}
        \label{fig:reso_NN}        
    \end{subfigure}
    \caption{Resolution of the reconstructed radius measured using the pseudo-event method from the three position reconstruction algorithms, shown in bins of the average PE between the pseudo-events and reconstructed radius of the original event.}
    \label{fig:reso_plot}
\end{figure}

\begin{figure}[h!tb]
    \centering
        \begin{subfigure}[b]{0.48\textwidth}
        \centering
        \includegraphics[scale=0.38]{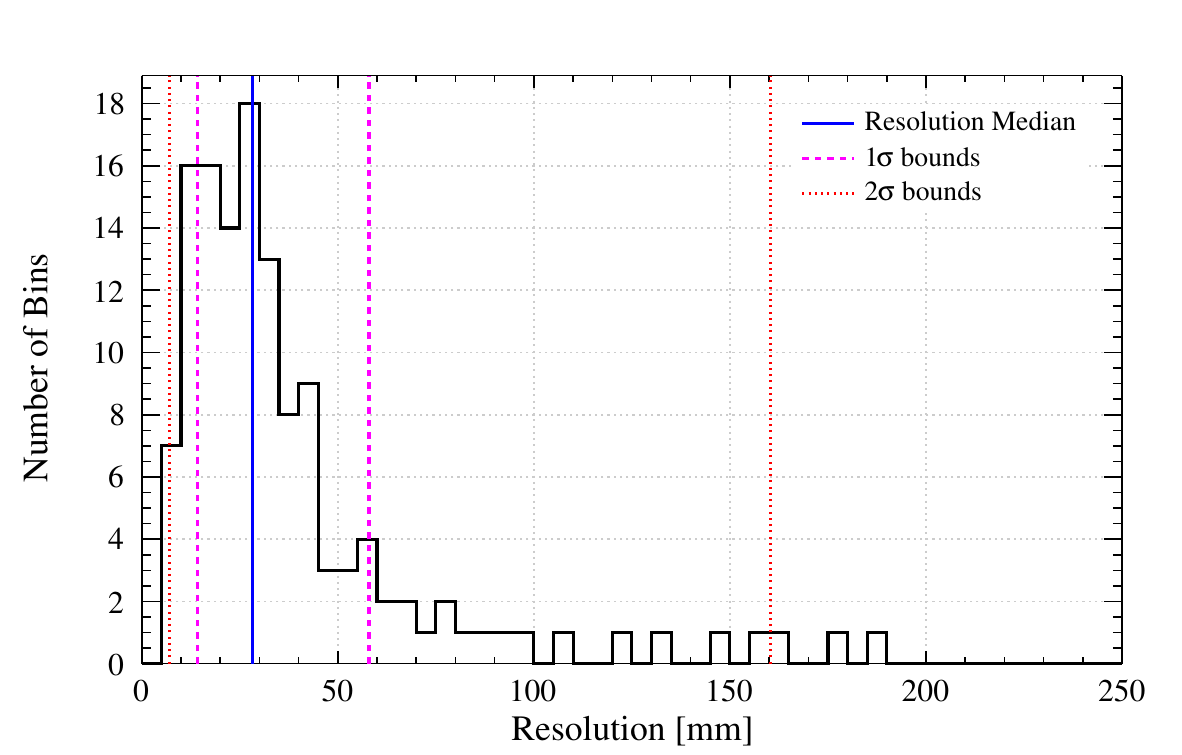}     
        \caption{Example 1D distribution - Likelihood PE-based}
        \label{fig:1D_reso_dist_with_lines}
    \end{subfigure}
    ~
    \begin{subfigure}[b]{0.48\textwidth}
        \centering
        \includegraphics[scale=0.4]{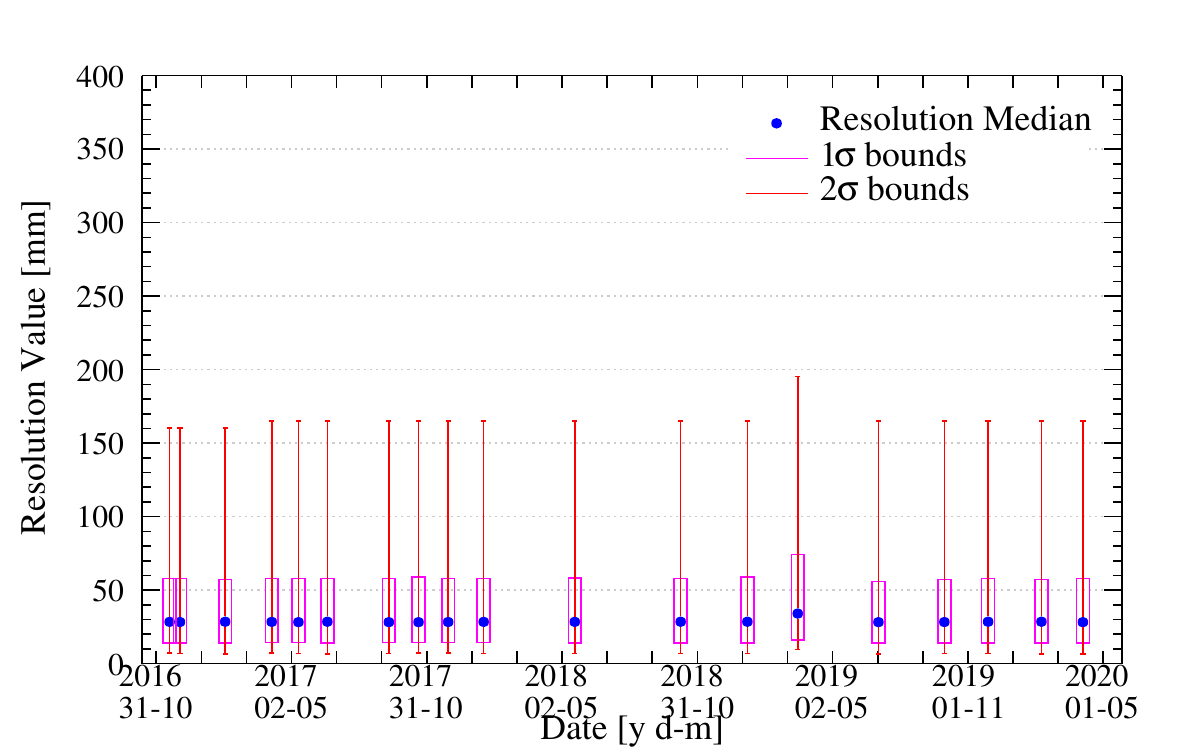}     
        \caption{Likelihood PE-based position  resolution}
        \label{fig:reso_MBL_time}
    \end{subfigure}
    ~
    \vspace{0.5pt}
    \begin{subfigure}[b]{0.48\textwidth}
        \centering
        \includegraphics[scale=0.4]{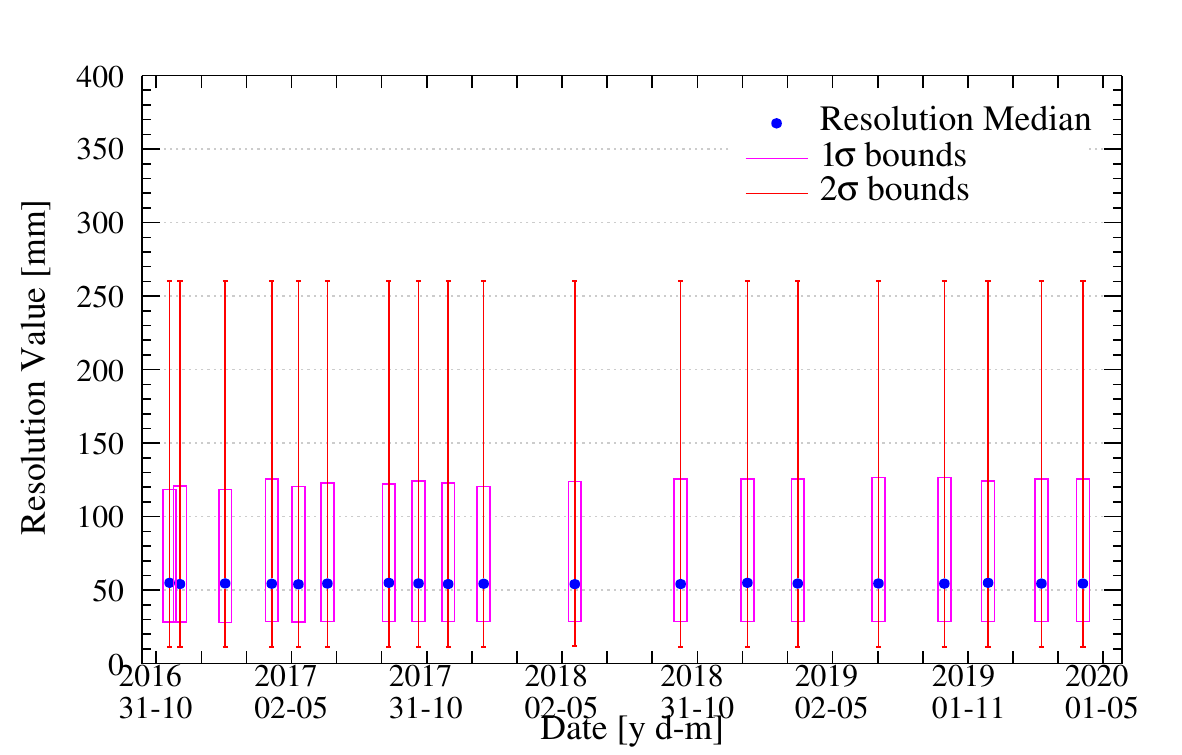}    
        \caption{Likelihood time-based position resolution}
        \label{fig:reso_TF2_time}        
    \end{subfigure}
    ~
    \begin{subfigure}[b]{0.48\textwidth}
        \centering
        \includegraphics[scale=0.4]{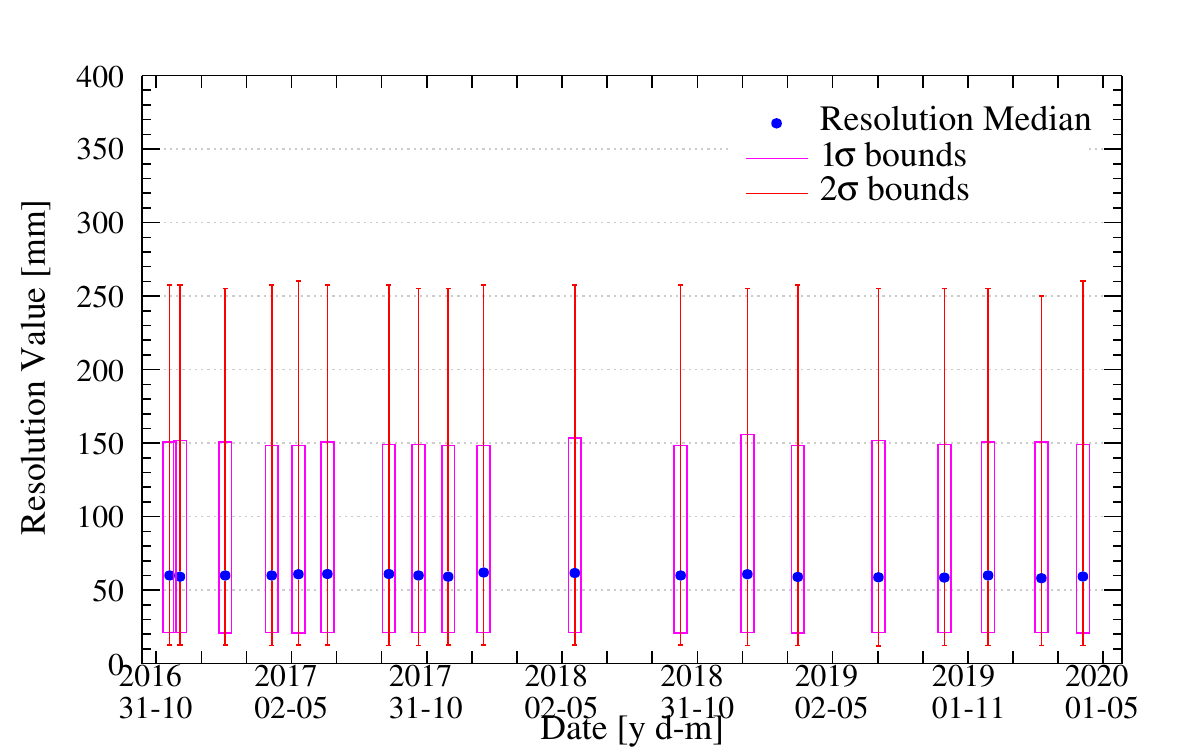}    
        \caption{Neural network PE-based position resolution}
        \label{fig:reso_NN_time}        
    \end{subfigure}
    \caption{In (a), the 2D resolution values per bin 
    were sorted onto a 1D histogram. Then from the distribution, the quantile points corresponding to the median, 1$\sigma$ (68\%) and 2$\sigma$ (95\%), are shown as a function of time in (b) through (d). These figures demonstrate the stability over time of the estimated resolution from the three position reconstruction algorithms.}
    \label{fig:reso_plot_timeseries}
\end{figure}

In the majority of cases, these distributions are described by the following function, which has a Gaussian core and an exponential component to account for non-Gaussian tails:
\begin{equation}
    g(\Delta x) = \left(\frac{1-\alpha}{\sigma\sqrt{2\pi}}\right) \, e^{ \scalebox{1.3}{-$\frac{(\Delta x-\mu)^2}{2\sigma^2}$} }  \,+\, \frac{\alpha}{2\tau} \, e^{ \scalebox{1.3}{-$\frac{\left|\Delta x-\mu\right|}{\tau}$} } \,,
    \label{eq:gaus+expo}
\end{equation}
where the $\alpha$ parameter varies between 0 (entirely Gaussian) and 1 (entirely exponential). 
As shown in Appendix~\ref{app:eff_sigma}, the characteristic width of this distribution is 
\begin{equation}
    \sigma_{\rm{eff}} =  \sqrt{(1-\alpha)\sigma^2 + 2\alpha\tau^2}
    \label{eq:resolution} \,.
\end{equation}

In the case of the FFNN reconstructed $z$ distributions at low $r$ and PE, a different fit function is required as shown in Fig.~\ref{fig:split_event_dist_NNZ_lowR}. 
The fit function chosen to fit this distribution is a sum of four normalized Gaussians:
\begin{equation}  
    g(\Delta z) = \left(\frac{w_{\rm{0}}}{\sigma_{\rm{0}}\sqrt{2\pi}}\right) \, e^{ \scalebox{1.3}{-$\frac{(\Delta z+\mu_{\rm{0}})^2}{2\sigma_{\rm{0}}^2}$}} + \left(\frac{w_{\rm{0}}}{ \sigma_{\rm{0}}\sqrt{2\pi}}\right) \, e^{ \scalebox{1.3}{-$\frac{(\Delta z-\mu_{\rm{0}})^2}{2\sigma_{\rm{0}}^2}$}} + \left(\frac{w_{\rm{1}}}{\sigma_{\rm{1}}\sqrt{2\pi}}\right) \, e^{ \scalebox{1.3}{-$\frac{(\Delta z)^2}{2\sigma_{\rm{1}}^2}$}} + \left(\frac{w_{\rm{2}}}{\sigma_{\rm{2}}\sqrt{2\pi}}\right) \, e^{ \scalebox{1.3}{-$\frac{(\Delta z)^2}{2\sigma_{\rm{2}}^2}$}} \,.
    \label{eq:4gaussian}
\end{equation}  
The distribution contains ``outer Gaussians'', which are symmetrically apart from the origin, and have the same scaling $w_{\rm{0}}$ and standard deviation $\sigma_{\rm{0}}$. 
They correspond to cases where the FFNN misclasifies only one of the two pseudo-events as originating from the neck, and the other from the AV, resulting in a large $\Delta z$.
The two central Gaussians with weights $w_{\rm{1}}$ and $w_{\rm{2}}$ have their means set to zero. 
As shown in Appendix~\ref{app:eff_sigma}, the characteristic width of this distribution is 
\begin{equation}
    \sigma_{\rm{eff}} =  \sqrt{2w_{\rm{0}} (\sigma_{\rm{0}}^2 + \mu_{\rm{0}}^2 ) + w_{\rm{1}} \sigma_{\rm{1}}^2 + w_{\rm{2}} \sigma_{\rm{2}}^2} .
    \label{eq:resolution_4gaus}
\end{equation}


The collection of plots in Figure~\ref{fig:resolution_delta_distributions} showcase some examples of the difference distributions of positions between the pseudo-events for each reconstruction algorithm. Distributions are shown for two regions of the detector: near the center and near the edges, for a chosen PE range of 100--200. For the likelihood algorithms, only $\Delta x$ is displayed as example, since $\Delta y$ and $\Delta z$ follow the same behavior. For the neural network, $\Delta x$ and $\Delta y$ share the same profile, but $\Delta z$ is different, so an additional pair of examples is provided for the latter.

Figure~\ref{fig:reso_plot} shows the position resolution of the three algorithms as a function of the average PE of the pseudo-events and the reconstructed radius of the original event. 
The position resolution improves as we get closer to the AV surface, and as the number of PE in the event increases.  

To quantify how the position reconstruction resolution evolves over time, for each 2D plot, 1D distributions of the resolution values were created. Each of the values were sorted into 100 bins, 5 mm wide each. Next, the locations of the median, 1$\sigma$ and 2$\sigma$ quantiles were extracted from the distribution. 
Shown in Figure~\ref{fig:1D_reso_dist_with_lines} is an example of the 1D distribution with lines added at the locations of these key quantiles.
The other parts of Figure~\ref{fig:reso_plot_timeseries} show the stability of these position resolution metrics across the second-fill operation of DEAP-3600 in 2016--2020.

\section{Summary}

The DEAP-3600 experiment developed three complementary algorithms for reconstructing the positions of interactions within LAr: 
a maximum likelihood algorithm based on PE patterns, a maximum likelihood algorithm utilizing time-of-flight information, and a machine-learning algorithm employing neural networks with PE patterns as input.
All three algorithms demonstrate comparable performance for position reconstruction within the LAr volume. However, in the detector's neck region, only the machine-learning approach provides reliable position resolution, as it was specifically designed and trained on a simulated dataset containing events from both the LAr volume and the neck region.

The successful identification and rejection of background events, particularly shadowed alpha decays occurring in the neck, 
is critical for enhancing the experiment's sensitivity to WIMP dark matter signals.
The two maximum-likelihood algorithms specialized for reconstructing the event position in the LAr volume were further tested
with data from a radioactive $^{22}$Na source, confirming their good agreement with simulations.

To further refine the position resolution determination, 
a fully data-driven method was developed.
The dependence of position resolution on both position and energy was studied in detail. 
This method demonstrated the excellent stability of the position resolution over the data collection period.
The capability of reconstructing the position of events in DEAP-3600 significantly enhances the detector's sensitivity 
to WIMP dark matter within the region of interest, improving its ability to detect rare nuclear recoils in the bulk LAr volume
while rejecting surface backgrounds.

\section{Acknowledgements}
We thank the Natural Sciences and Engineering Research Council of Canada (NSERC),
the Canada Foundation for Innovation (CFI),
the Ontario Ministry of Research and Innovation (MRI), 
and Alberta Advanced Education and Technology (ASRIP),
the University of Alberta,
Carleton University, 
Queen's University,
the Canada First Research Excellence Fund through the Arthur B.~McDonald Canadian Astroparticle Physics Research Institute,
Consejo Nacional de Ciencia y Tecnolog\'ia Project No. CONACYT CB-2017-2018/A1-S-8960, 
DGAPA UNAM Grants No. PAPIIT IN108020 and IN105923, 
and Fundaci\'on Marcos Moshinsky,
the European Research Council Project (ERC StG 279980),
the UK Science and Technology Facilities Council (STFC) (ST/K002570/1 and ST/R002908/1),
the Leverhulme Trust (ECF-20130496),
the Russian Science Foundation (Grant No. 21-72-10065),
the Spanish Ministry of Science and Innovation (PID2019-109374GB-I00) and the Community of Madrid (2018-T2/ TIC-10494), 
the International Research Agenda Programme AstroCeNT (MAB/2018/7) funded by the Foundation for Polish Science (FNP) from the European Regional Development Fund,
and the Polish National Science Centre (2022/47/B/ST2/02015).
Studentship support from
the Rutherford Appleton Laboratory Particle Physics Division,
STFC and SEPNet PhD is acknowledged.
We thank SNOLAB and its staff for support through underground space, logistical, and technical services.
SNOLAB operations are supported by the CFI
and Province of Ontario MRI,
with underground access provided by Vale at the Creighton mine site.
We thank Vale for their continuing support, including the work of shipping the acrylic vessel underground.
We gratefully acknowledge the support of the Digital Research Alliance of Canada,
Calcul Qu\'ebec,
the Centre for Advanced Computing at Queen's University,
and the Computational Centre for Particle and Astrophysics (C2PAP) at the Leibniz Supercomputer Centre (LRZ)
for providing the computing resources required to undertake this work.

\appendix
\section{Data-Driven Estimation Method for Position Resolution}
\label{app:eff_sigma}
The spherical radius $r$ expressed in terms of three Cartesian coordinates $x,y,z$ is 
\begin{equation}
    r(x,y,z) = \sqrt{x^2 \,+\, y^2 \,+\, z^2} \quad.
    \label{eq:Radius}
\end{equation}
From standard error propagation, assuming uncorrelated variables, the error on the radius is
\begin{equation}
    \delta r = \frac{\sqrt{ (x\delta x)^2 + (y\delta y)^2 + (z\delta z)^2}}{r} \quad.
\end{equation}

Under the assumption of spherical symmetry, we assume that $\delta x$, $\delta y$ and $\delta z$ are constant over fixed spherical coordinates $\theta$ and $\phi$, thus the average resolution $\overline{\delta r}$ of the reconstructed radius can be obtained by integrating over the solid angle: 
\begin{equation}
    \overline{\delta r} = \frac{1}{4\pi} \int_{\phi=0}^{2\pi} \int_{\theta=0}^{\pi} \sqrt{(\sin\theta \cos\phi\ \delta x)^2 \,+\, 
    (\sin\theta \sin\phi\ \delta y)^2 \,+\, (\cos\theta\ \delta z)^2} \, 
    \sin\theta \, d\theta\, d\phi \quad.
\end{equation}
Performing the integral, 
\begin{equation}
    \overline{\delta r} = \sqrt{\frac{ (\delta x)^2 + (\delta y)^2 + (\delta z)^2 }{3}} \quad.\label{eq:r_reso_final}
\end{equation}

The errors on the Cartesian reconstructed positions are determined from the characteristic widths of the distributions of the distances between pairs of pseudo-events, $\Delta x$, $\Delta y$ and $\Delta z$ as explained in Section~\ref{sec:resolution}. These characteristic widths are derived from the variance of the distribution from the fit function used in each case.

First, expressing the mixture distribution of a Gaussian with exponential tails from Eq.~\ref{eq:gaus+expo} as:
\begin{equation}
    g(\Delta x) = (1-\alpha)f_1(\Delta x)+\alpha f_2(\Delta x) \,,
\end{equation}
where $f_{\rm{1}}$ and $f_{\rm{2}}$ are respectively the normalized Gaussian and exponential tail distributions, 
%
the expectation value and the variance of $g(\Delta x)$ are respectively:
\begin{eqnarray}
    {\rm E}[g(\Delta x)] &= & {\rm E}[ (1-\alpha)f_1(\Delta x)+\alpha f_2(\Delta x)] \nonumber\\
     &= & (1-\alpha)\mu + \alpha \mu = \mu \,, \\
    {\rm V}[g(\Delta x)] &= & (1-\alpha)\int x^2f_1(x) \,dx\ + \alpha\int x^2f_2(x) \,dx\ -\mu^2 \nonumber\\
     &= & (1-\alpha)(\sigma^2 + \mu^2) + \alpha(2\tau^2 + \mu^2) -\mu^2 \nonumber\\
     &= & (1-\alpha)\sigma^2 + 2\alpha\tau^2 \,.
\end{eqnarray}
The characteristic width of the Gaussian-exponential mixture distribution is therefore:
\begin{equation}
    \label{appendix_eq:gaus+expo}
    \sigma_{\rm{eff}} = \sqrt{(1-\alpha)\sigma^2 + 2\alpha\tau^2} \,.
\end{equation}

In the case where the mixture distribution is a sum of four Gaussians, a similar process is followed. Starting with Eq.~\ref{eq:4gaussian},
the expectation value and the variance are:
\begin{eqnarray}
    {\rm E}[g(\Delta x)] &= & \sum_i w_i \mu_i = w_0 (-\mu_0 + \mu_0) = 0 \,, \\
    {\rm V}[g(\Delta x)] &= & \sum_i w_i (\sigma_i^2 + \mu_i^2) - {\rm E}[g(\Delta x)]^2 \nonumber\\
     &= & 2w_0 (\sigma_0^2 + \mu_0^2 ) + w_1 \sigma_1^2 + w_2 \sigma_2^2 \,.
\end{eqnarray}
Therefore, in this case the characteristic width is:
\begin{eqnarray}
    \label{appendix_eq:4gaussian}
    \sigma_{\rm{eff}} &= & \sqrt{2w_0 (\sigma_0^2 + \mu_0^2 ) + w_1 \sigma_1^2 + w_2 \sigma_2^2}
\end{eqnarray}

Based on Eq.~\ref{appendix_eq:gaus+expo} or Eq.~\ref{appendix_eq:4gaussian}, the appropriate characteristic widths for $\delta x, \delta y, \delta z$ are substituted into Eq.~\ref{eq:r_reso_final}, which becomes: 
\begin{equation}
    \overline{\delta r} = \sqrt{\frac{ (\sigma_{\rm{eff}}^x)^2 + (\sigma_{\rm{eff}}^y)^2 + (\sigma_{\rm{eff}}^z)^2 }{3}} . \label{eq:r_reso_final_1}
\end{equation}

An additional correction needs to be taken into account to obtain a true estimate of the position resolution.  
Mathematically, the true resolution is defined as the width of the distribution $P\left( \left| \Vec{x}_{\rm rec} - \Vec{x}_{\rm true} \right| \right)$ of the difference between the true ($\Vec{x}_{\rm true}$) and reconstructed ($\Vec{x}_{\rm rec}$) position. 
In actual data, two pseudo-events are generated by randomly selecting PE units from the original event. 
Position reconstruction of two pseudo-events provides two fit positions ($\Vec{x}_1$ and $\Vec{x}_2$), and the distribution of their difference is $Q\left( \left| \Vec{x}_{1} - \Vec{x}_{2} \right| \right)$ (as in Figure~\ref{fig:split_event_dist_mbl_lowR}).
So far, the distribution $Q$, which is different from the distribution $P$, is used to calculate the position resolution. 
If the distribution $Q$ follows Eq.~\ref{appendix_eq:gaus+expo} or Eq.~\ref{appendix_eq:4gaussian} then we can assume the distribution $P$ follows it as well, as long as photon emission and detection are uncorrelated.
The following equation shows the relationship between the $P$ and $Q$ distributions:
\begin{equation}
    Q\left( \left| \Vec{x}_{1} - \Vec{x}_{2} \right| \right) = \int_{V} d^3 \Vec{x}_{\rm true} P\left( \left| \Vec{x}_{1} - \Vec{x}_{\rm true} \right| \right) P\left( \left| \Vec{x}_{2} - \Vec{x}_{\rm true} \right| \right) \,.
    \label{eq:P_and_Q_distn}
\end{equation}

\begin{figure}[htbp]
    \centering
    \includegraphics[scale=0.5]{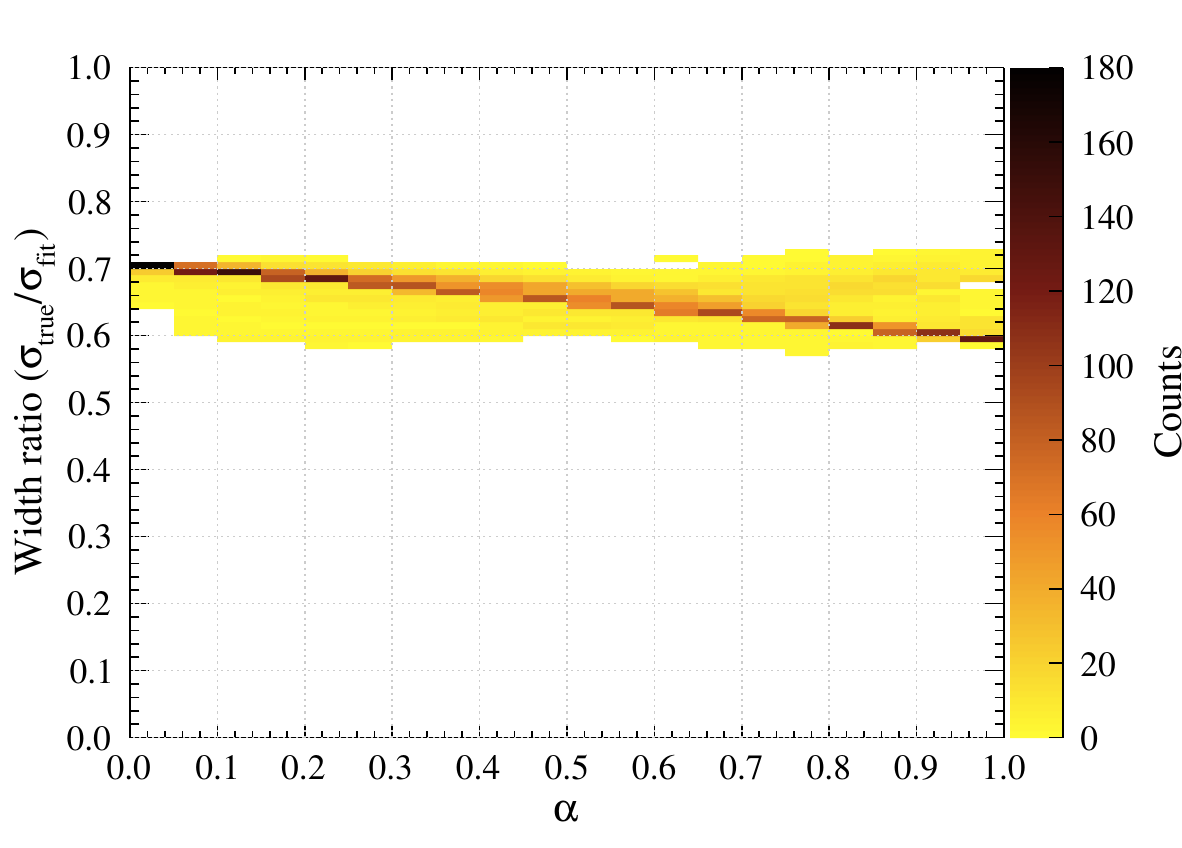}
    \caption{Ratio of $\sigma_{\rm{true}}$ to $\sigma_{\rm{fit}}$ versus the $\alpha$ parameter in the toy MC.}
    \label{fig:sigratio_alpha}
\end{figure}

For distributions where only Eq.~\ref{appendix_eq:gaus+expo} is used, the relationship between the ratio of $P-$resolution ($\sigma_{\rm true}$) to $Q-$resolution ($\sigma_{\rm rec}$) and the $\alpha$ parameter is obtained using a toy MC. First, $\sigma_{\rm true}$ is obtained from the parameters $\alpha,\,\sigma$, and $\tau$, varied in a range comparable to the observed data. Then, two random events are generated following Eq.~\ref{appendix_eq:gaus+expo} and afterwards fitted to obtain the effective resolution $\sigma_{\rm fit}$. The simulation results are shown in Fig.~\ref{fig:sigratio_alpha}.
A correction factor of 2/3, obtained when setting $\alpha = 0.5$ in the linear $\alpha$-based correction, was therefore used to scale the data-driven $Q-$resolution values calculated with Eq.~\ref{eq:r_reso_final_1} in order to obtain the $P$-resolution values shown in Fig.~\ref{fig:reso_plot} and Fig.~\ref{fig:reso_plot_timeseries}.  
The same correction factor was applied to cases involving the four-Gaussian mixture distribution for neural network position resolution calculations.


\bibliographystyle{deap}
\bibliography{Bibliography.bib}

\end{document}